\documentclass[11pt,a4paper]{article}

\def\PWG{{\tt{POWHEG}}}
\def\PYT{{\tt{PYTHIA 8}}}
\usepackage{bold-extra}

\usepackage{graphicx}

\graphicspath{ {./Figures/} }

\usepackage{authblk}
\usepackage{comment}

\pdfoutput=1
\usepackage{color}
\usepackage{todonotes}
\usepackage{subfigure}
\usepackage{booktabs}
\usepackage{array}
\usepackage{multirow,rotating}
\usepackage{amsmath,amssymb,bm,cite,graphicx,wasysym,mathrsfs,dsfont}
\usepackage{slashed,relsize}
\usepackage[colorlinks=true,linkcolor=blue,anchorcolor=black,citecolor=blue,filecolor=cyan,menucolor=red,runcolor=filecolor,urlcolor=blue,bookmarks=true,bookmarksnumbered=true]{hyperref}

\usepackage[font=small,labelfont=bf]{caption}

\DeclareMathOperator{\arctanh}{arctanh}

\def\beq{\begin{equation}\displaystyle\displaystyle}
\def\eeq{\end{equation}}
\def\bea{\begin{eqnarray}\displaystyle} 
\def\eea{\end{eqnarray}}

\def\({\left(}
\def\){\right)}
\def\bry{\begin{array}}
\def\ery{\end{array}}

\title{\bf High-energy EFT probes with fully differential \\ Drell--Yan measurements \vspace{0.4cm}} 
\date{}

\author[1,2]{Giuliano Panico}

\author[3]{Lorenzo Ricci}

\author[3,4,5]{Andrea Wulzer}

\affil[1]{\emph{Dipartimento di Fisica e Astronomia, Universit\`a di Firenze, Sesto Fiorentino, Italy}} 
\affil[2]{\emph{INFN, Sezione di Firenze, Sesto Fiorentino, Italy}}
\affil[3]{\emph{Institut de Th\'eorie des Ph\'enomenes Physiques, EPFL, Lausanne, Switzerland}}
\affil[4]{\emph{Theoretical Physics Department, CERN, Geneva, Switzerland}}
\affil[5]{\emph{Dipartimento di Fisica e Astronomia, Universit\'a di Padova, Italy}}

\begin{document}
\baselineskip=14pt

\maketitle

\begin{abstract}
We study the potential of fully-differential measurements of high-energy dilepton cross-sections at the LHC to probe heavy new physics encapsulated in dimension-$6$ interaction operators. The assessment is performed in the seven-dimensional parameter space of operators that induce energy-growing corrections to the Standard Model partonic cross-sections at the interference level, and in the two-dimensional subspace associated with the W and Y parameters. A considerable sensitivity improvement is found relative to single-differential measurements, owing to the possibility of probing at the interference level more directions in the seven-dimensional parameter space. The reduction of parton distribution function uncertainties in the fully-differential fit is also found to play a significant role. The results are interpreted in the minimal $Z^\prime$ new-physics model, providing a concrete illustration of the advantages of the fully-differential analysis. We find that high-energy dilepton measurements can extend the $Z^\prime$ exclusion and discovery potential well beyond the reach of direct searches in a large region of the parameter space.
\end{abstract}

\newpage

\begingroup
\tableofcontents
\endgroup 

\setcounter{equation}{0}
\setcounter{footnote}{0}
\setcounter{page}{1}

\newpage

\section{Introduction}

The Drell--Yan (DY) production of high-energy lepton pairs at the LHC and at its High-Luminosity successor (HL-LHC) is a powerful probe of new physics encapsulated in Effective Field Theory (EFT) operators~\cite{deBlas:2013qqa,Farina:2016rws,Dawson:2018dxp,Ricci:2020xre}. It is particularly sensitive to the operators that induce growing-with-energy effects in the partonic cross-sections through their interference with the Standard Model (SM) contribution. These are the flavor-universal operators constructed with one quark and one lepton current, listed in Table~\ref{tab:GenOp}.\footnote{Operators that are diagonal in the fermions mass-basis, but not flavor-universal, can also interfere. However they are generically too strongly constrained by flavor measurements to be relevant.} Thanks to the high energy that is available at the LHC, DY measurements can extend the current experimental sensitivity to these operators by more than one order of magnitude. This generically holds also for operator combinations that are already accurately probed by the ElectroWeak Precision Tests (EWPT) performed at LEP and other low-energy experiments~\cite{Farina:2016rws}. 

The relevance of high-energy DY measurements actually extends beyond their intrinsic sensitivity to new physics. The DY processes are the simplest concrete incarnation of the so-called ``high-$p_T$'' probes of new physics, aimed at capturing growing-with-energy EFT effects in the tail of kinematical distributions. Other candidate processes include diboson and boson-plus-Higgs~\cite{Butter:2016cvz,Zhang:2016zsp,Liu:2018pkg,Green:2016trm,Biekoetter:2014jwa,Falkowski:2015jaa,Baglio:2017bfe,Franceschini:2017xkh,Panico:2017frx,Grojean:2018dqj,Banerjee:2018bio}, diquark and top production~\cite{Alioli:2017nzr,Farina:2018lqo} and others~\cite{Greljo:2017vvb,Henning:2018kys}. Robust phenomenological studies of DY, possibly accompanied by preliminary LHC experimental results and HL-LHC projections, are instrumental to reveal issues and outline solutions towards the development of the high-$p_T$ measurement program and its extension to the other processes listed above. For instance the DY process is the simplest playground for the assessment of the impact of Parton Distribution Function (PDF) uncertainties and of the potential advantages for new physics sensitivity of reducing them with a simultaneous fit to the EFT and to the PDF~\cite{Carrazza:2019sec}.

The DY final state is so simple that it can be characterized completely with fully-differential cross-section measurements. Fully (i.e.~triply) differential measurements of the neutral DY process have already been performed by ATLAS with early run-$1$ data~\cite{Aaboud:2017ffb}.\footnote{Double differential measurements have been performed in Refs.~\cite{Aad:2016zzw,CMS:2014hga}.} Quantifying the sensitivity to the EFT operators of fully-differential measurements of the neutral ($\ell^+\ell^-$) and charged ($\ell\nu$) DY processes at the LHC, and its improvement relative to the sensitivity of the single-differential analysis, is the purpose of the present paper. We employ the framework of Ref.~\cite{Ricci:2020xre} to obtain accurate DY predictions in the presence of the $7$ EFT operators of Table~\ref{tab:GenOp} by analytic event reweighting. This allows us to explore the $7$-dimensional EFT parameter space efficiently, based on few Monte Carlo simulations performed under the SM hypothesis. The large dimensionality of the EFT parameter space, combined with the large number of analysis bins that are employed for the fully-differential analysis, would have made this project computationally too demanding if employing a Monte Carlo implementation not based on reweighting. The statistical treatment of uncertainties is also borrowed from Ref.~\cite{Ricci:2020xre}.

The paper is organized as follows. In Section~\ref{treedits} we develop a qualitative understanding of the fully-differential measurement potential to improve the sensitivity to new physics in several directions of the $7$-dimensional EFT parameter space. Sensitivity projections for HL-LHC are presented in Section~\ref{WYSubSec} focusing on the $2$-dimensional ``Universal'' subspace that corresponds to the W any Y parameters~\cite{Farina:2016rws,Ricci:2020xre}. The $7$-dimensional global analysis results are presented in Section~\ref{GCCSubSec} by employing a basis in the parameter space that is designed to outline the hierarchy in sensitivity among the various directions and to make the differential analysis advantages manifest. In Section~\ref{sec:Zprime} we quantify the fully-differential analysis potential concretely in a simple Beyond the SM (BSM) Z$^\mathbf{\prime}$ scenario. We report our conclusions in Section~\ref{sec:conc}. Finally, in the appendices we discuss the definition of the kinematical variables beyond the tree-level approximation (Appendix~\ref{app:Kin}), we give the explicit expressions for the tree-level amplitude (Appendix~\ref{app:Form}) and we report some selected results of our analysis for the LHC run 3 (Appendix~\ref{app:LHCProj})

\begin{table}
\centering
\begin{tabular}{l}
current-current quark-lepton operators\\
\hline
\hspace{1em}\rule{0pt}{1.5em}${\cal O}_{lq}^{(3)}=(\bar{l}_L\sigma_I \gamma^\mu {l}_L)(\bar{q}_L\sigma_I\gamma_\mu {q}_L)$\\
\hspace{1em}\rule{0pt}{1.5em}${\cal O}_{lq}^{(1)}=(\bar{l}_L\gamma^\mu {l}_L) (\bar{q}_L\gamma_\mu {q}_L)$\\
\hspace{1em}\rule{0pt}{1.5em}${\cal O}_{eu}=(\bar{e}_R\gamma^\mu {e}_R) (\bar{u}_R\gamma_\mu {u}_R)$\\
\hspace{1em}\rule{0pt}{1.5em}${\cal O}_{ed}=(\bar{e}_R\gamma^\mu {e}_R) (\bar{d}_R\gamma_\mu {d}_R)$\\
\hspace{1em}\rule{0pt}{1.5em}${\cal O}_{lu}=(\bar{l}_L \gamma^\mu {l}_L) (\bar{u}_R \gamma_\mu {u}_R)$\\
\hspace{1em}\rule{0pt}{1.5em}${\cal O}_{ld}\hspace{0.11cm}=(\bar{l}_L \gamma^\mu {l}_L) (\bar{d}_R \gamma_\mu {d}_R)$\\
\hspace{1em}\rule{0pt}{1.5em}${\cal O}_{qe}=(\bar{q}_L \gamma^\mu {q}_L) (\bar{e}_R \gamma_\mu {e}_R)$
\end{tabular}
\hspace{2em}
\begin{tabular}{ll}
\multicolumn{2}{c}{W and Y current-current operators}\\
\hline
\hspace{1em}\rule{0pt}{1.5em}${\cal O}'_{2W}=J_L^{a,\mu} J^a_{L,\mu}$ & $J_L^{a,\mu} = \sum_f \bar f \gamma^\mu T^a f$\\
\hspace{1em}\rule{0pt}{1.5em}${\cal O}'_{2B} = J_Y^{\mu} J_{Y,\mu}$ & $J_Y^{\mu} = \sum_f \bar f \gamma^\mu Y_f f$\\
\hspace{1em}\rule{0pt}{3.0em}$G_{Lq}^{(3)} = \frac{1}{2} G'_{2W}$&\\
\hspace{1em}\rule{0pt}{1.5em}$G_{Lq}^{(1)} = -\frac{1}{6} G'_{2B}$ & $G_{eu} = -\frac{4}{3} G'_{2B}$\\
\hspace{1em}\rule{0pt}{1.5em}$G_{ed} = \frac{2}{3} G'_{2B}$ & $G_{lu} = - \frac{2}{3} G'_{2B}$\\
\hspace{1em}\rule[-.7em]{0pt}{2.2em}$G_{ld} = \frac{1}{3} G'_{2B}$ & $G_{qe} = - \frac{1}{3} G'_{2B}$
\end{tabular}
\caption{Left table: Quark-lepton current-current operators in the Warsaw basis~\cite{Grzadkowski:2010es}. The operators are flavor universal  and thus the generation indices have been suppressed. Right table: The operators related to the W and Y parameters, together with their rewriting in terms of Warsaw basis operators.
\label{tab:GenOp}} 
\end{table}

\section{Fully-differential Drell-Yan}\label{sec:diff_distr}

We start our investigation of the fully-differential DY dilepton production and its sensitivity to new physics by developing a semi-analytic qualitative understanding based on the structure of the tree-level distributions. Quantitative estimates of the sensitivity are performed in Sections~\ref{WYSubSec} and~\ref{GCCSubSec}.

\subsection{Tree-level distributions}\label{treedits}

Consider the neutral process $q\bar q \to \ell^+ \ell^-$. The fully-differential cross-section is given by
\begin{align}\label{eq:sigma_ll}
\begin{aligned}
\frac{d^3\sigma}{d m_{\ell\ell}^2\, dc_*\, dy}=\frac{\tau}{3\cdot 64\,\pi\, m_{\ell\ell}^4} \sum_{q} \Big\{\left[ (1+c_*)^2 \mathcal{L}_q(\tau,y)
+(1-c_*)^2 \mathcal{L}_q(\tau,-y)\right]P_s^q (m_{\ell\ell})\qquad \\
+ \left[(1-c_*)^2 \mathcal{L}_q(\tau,y)+(1+c_*)^2 \mathcal{L}_q(\tau,-y)
\right]P_o^q(m_{\ell\ell}) \Big\}\,,
\end{aligned}
\end{align}
where $m_{\ell\ell}=\sqrt{\hat{s}}$ is the dilepton invariant mass and $\tau=\hat{s}/S$ (with $\sqrt{S}$ the collider energy), while $y$ is the absolute value of the rapidity (relative to the beam axis) of the dilepton system. We define $c_*=\cos\theta_*$ as the cosine of the angle formed, in the rest frame of the dilepton pair, by the charge-minus lepton and the direction of motion of the dilepton rest frame relative to the lab frame. At tree-level, $\theta_*$ as defined above is the angle between the $\ell^-$ and the most energetic incoming parton.\footnote{It is essential \emph{not} to define $\theta_*$ with respect to a fixed beam-axis orientation. With that definition, the fully-differential cross-section in eq.~(\ref{eq:sigma_ll}) would depend only on the combination $(P_s^q +P_o^q)$ like the single-differential cross-section in eq.~(\ref{eq:sigma_s}), and all the advantages of the fully-differential analysis would be lost.\label{foot1}} The detailed definition of the kinematical variables beyond tree-level is reported in Appendix~\ref{app:Kin}. 

The sum in eq.~(\ref{eq:sigma_ll}) spans over the light quarks $q=\{u,d,c,s,b\}$, and, for each quark species $q$, ${\mathcal{L}}_q$ is the product of the corresponding $q$ and $\overline{q}$ parton distribution functions (PDFs), namely
\begin{equation}\displaystyle\label{pdf}
{\mathcal{L}}_q(\tau,y)=f_q(\sqrt{\tau}\, e^y;m_{\ell\ell}^2)\,f_{\overline{q}}(\sqrt{\tau} \,e^{-y};m_{\ell\ell}^2)\,.
\end{equation}

The coefficient functions $P_s^q$ ($P_o^q$) parametrize the contributions, including both SM and new physics, from the subprocesses where the chirality of the incoming quarks is the same (opposite) one of the outgoing leptons. Our target new physics operators are flavor-universal, like the SM contribution to the scattering amplitudes. Therefore the coefficient functions are the same for all the up-type and for all the down-type quarks, for a total of four independent functions $P_s^{u,d}$ and $P_o^{u,d}$. In the high energy regime $m_{\ell\ell}\gg m_Z$, and at the linear interference level in the new physics contribution, the coefficient functions read
\begin{equation}\displaystyle\label{P}
\begin{array}{l}
    P_s^{u,d} (m_{\ell\ell}) \simeq P_{{\rm{SM}},s}^{u,d} +m_{\ell\ell}^2 \left( \vec{V}_s^{u,d} \cdot \vec{G} \right) \equiv P_{{\rm{SM}},s}^{u,d} + m_{\ell\ell}^2 \,G_s^{u,d}   \,,\\
    \rule{0pt}{2.em}P_s^{u,d} (m_{\ell\ell}) \simeq P_{{\rm{SM}},o}^{u,d} + m_{\ell\ell}^2 \left( \vec{V}_o^{u,d} \cdot \vec{G} \right) \equiv P_{{\rm{SM}},s}^{u,d} + m_{\ell\ell}^2 \,G_o^{u,d}   \,,
\end{array}
\end{equation}
where $\vec G$ denotes the Wilson coefficients of the seven effective four-fermion operators defined in Table~\ref{tab:GenOp}. In these expressions both the SM terms $P_{{\rm{SM}},s}^{u,d}$ and $P_{{\rm{SM}},o}^q$, and the vectors $\vec{V}_{s,o}^{u,d}$, are kinematics-independent numerical coefficients, reported in Appendix~\ref{app:Form}. At the linear level, and up to tiny effects suppressed by $m_Z^2/m_{\ell\ell}^2$, the neutral DY cross-section depends on new physics only through the  four linear combinations $\smash{{G}^{u,d}_{s,o}=\vec{V}_s^{u,d} \cdot \vec{G}}$ of the seven Wilson coefficients. With the fully-differential analysis we can probe each of these four directions in the new physics parameters space independently, at least in line of principle.

Consider for comparison the single-differential cross-section $d \sigma/d m_{\ell\ell}^2$. By integrating eq.~(\ref{eq:sigma_ll}) over $c_*$ and $y$, we get
\begin{equation}\displaystyle\label{eq:sigma_s}
\frac{d\sigma}{d m_{\ell\ell}^2} = \frac{\tau}{72\, \pi\, m_{\ell\ell}^4} \sum_q \frac{d {\cal L}_{q}}{d \tau} \left[P_s^q(m_{\ell\ell}) + P_o^q(m_{\ell\ell})\right]\,,
\end{equation}
where the parton luminosities are defined as
\begin{equation}\displaystyle
\frac{d {\cal L}_{q}}{d \tau} = \int\limits_{-y_{\rm{max}}}^{y_{\rm{max}}} \!\!\! dy\, {\cal L}_q(\tau, y)\,,\;\;\;\;\;y_{\rm{max}} = - {\frac{1}{2} \log \tau}\,.
\end{equation}
We see that the single-differential cross-section only depends on the sum of the ``$s$'' and ``$o$'' coefficient functions. At linear level, using eq.~(\ref{P}), it is thus only sensitive to ${G}^{u}_{s}+{G}^{u}_{o}$ and ${G}^{d}_{s}+{G}^{d}_{o}$, i.e. to two combinations of the four directions in the EFT parameter space that the fully-differential analysis can probe. Actually it is not difficult to see that the single-differential analysis is not even sensitive to ${G}^{u}_{s}+{G}^{u}_{o}$ and ${G}^{d}_{s}+{G}^{d}_{o}$ independently, but only to the combination
\begin{equation}\displaystyle\label{singdir}
 (G_o^u + G_s^u) + 0.5\,(G_o^d + G_s^d)\,.
\end{equation}
This is because the ratio between the up and the down quarks luminosities (that dominate over the one of the other quark flavors) is nearly constant in $\tau$ in the most sensitive energy range $m_{\ell\ell}\sim 1 - 2$~TeV. The ratio is approximately equal to $2$ owing to the valence quarks content of the proton.

The advantages of performing a fully-differential measurement can now be appreciated by analyzing the various regions in the $(c_*,y)$ kinematic space. In the kinematical regime with small center of mass rapidity ($y \simeq 0$), one has ${\cal L}_q(\tau, y) \simeq {\cal L}_q(\tau, -y)$, and the cross-section in eq.~(\ref{eq:sigma_ll}) becomes proportional to  $(1+c_*^2) {\cal L}_q(\tau, 0)(P_s^q + P_o^q)$. Hence this region provides sensitivity to the same combination of Wilson coefficients that can be probed through the
$d \sigma/d m_{\ell\ell}^2$ distribution. On the other hand, in the region with large $y$ we have ${\cal L}_q(\tau, y) \gg {\cal L}_q(\tau, -y)$ for the (dominant) up and down quarks because the valence quarks are typically more energetic than the sea anti-quarks. Therefore the cross-section is proportional to $(1+c_*)^2 P_s^q + (1-c_*)^2 P_o^q$ and it is sensitive to both $P_s^q$ and $P_o^q$ for $c_* \simeq +1$ and $c_* \simeq -1$, respectively.

Measuring the fully-differential distribution can also mitigate the degeneracy between the up and down quark contributions that is due, as previously discussed, to the similar shape of the parton luminosities. Indeed the dependence of ${\cal L}_q(\tau, y)$ on $y$ is significantly different for the two quark species. In particular the up quark distribution is peaked at larger values of $y$ than the one of the down quark.\footnote{A detailed discussion of this feature can be found in Ref.~\cite{Panico:2016ary} (see in particular Figure~3).} 

The  discussion above shows that the fully-differential cross-section measurement has the potential to disentangle the four $G_{s,o}^{u,d}$ linear combinations of Wilson coefficients. This is a significant improvement relative to the single-differential measurement that is sensitive to one combination only. The quantitative assessment of this improvement is postponed to Section~\ref{GCCSubSec}. 

It should be stressed that our findings are based on the dependence of the cross-section on the Wilson coefficients at the linear order. At the quadratic level, all Wilson coefficients enter in the $P_{s,o}^q$ functions with comparable coefficients (see the explicit expression for the amplitudes in Appendix~\ref{app:Form}). Therefore also the combinations of parameters that do not enter or are suppressed in the linear term can be determined through their quadratic contributions. These combinations are all expected to be tested less effectively than the ones contributing to the linear terms, but with similar precision among them.

A similar analysis can be performed for the charged DY process $q\bar q' \to \ell \nu$. In this case, however, a fully-differential measurement has a milder impact. The reason for this is twofold. First, the charged process is only affected by one operator, namely ${\mathcal{O}}_{lq}^{(3)}$, so that no issue in disentangling various new physics contributions is present. Second, due to the presence of a neutrino, only two independent kinematic variables can be accessed, for instance the transverse momentum of the charged lepton $p_{T,\ell}$ and its rapidity $\eta_\ell$. The new-physics contributions depend on the center of mass energy $\sqrt{\hat s}$, which is closely correlated with $p_{T,\ell}$, but has a very mild correlation with $\eta_\ell$. The additional benefit of considering both kinematical variables rather than only $p_{T,\ell}$ is therefore expected to be small. However it should be taken into account that more differential information in the charged channel might help reducing the impact of PDF uncertainties in the combination with the fully-differential neutral DY measurements. Indeed some advantage of the doubly-differential measurement in charged DY will be observed in the analyses presented below.

\subsection{Bounds on the Universal parameters W and Y}\label{WYSubSec}

As a first quantitative analysis we focus on the specific set of dimension-6 operators related to the Universal parameters W and Y. As in Refs.~\cite{Farina:2016rws,Ricci:2020xre}, we define the W and Y parameters
\begin{equation}\displaystyle
G'_{2W} = - \frac{g^2}{2 m_W^2} \textrm{W}\,, \qquad \quad G'_{2B} = - \frac{g'^2}{2 m_W^2} \textrm{Y}\,,
\end{equation}
 in terms of the coefficients of the four-fermion operators
\begin{equation}\displaystyle
{\cal O}'_{2W} = J_L^{a,\mu} J^a_{L,\mu}\,,\qquad {\cal O}'_{2B} = J_Y^{\mu} J_{Y,\mu}\,,
\end{equation}
built from the $\textrm{SU}(2)_L$ and hypercharge $\textrm{U}(1)_Y$ currents
\begin{equation}\displaystyle
J_L^{a,\mu} = \sum_f \bar f \gamma^\mu T^a f\,, \qquad J_Y^{\mu} = \sum_f \bar f \gamma^\mu Y_f f\,.
\end{equation}
In the DY processes, ${\cal O}'_{2W}$ and ${\cal O}'_{2B}$ are equivalent to lepton-quark operators.\footnote{This only holds up to EW running effects. These effects are included in our prediction as in Ref.~\cite{Ricci:2020xre}, duly taking into account the contribution to the running of the quark-quark and lepton-lepton components of ${\cal O}'_{2W}$ and ${\cal O}'_{2B}$. The operator coefficients, and in turn W and Y, are defined at the renormalization scale of $10$~TeV.\label{runfoot}} The explicit correspondence is given in Table~\ref{tab:GenOp}. 

There are some crucial differences between the W and the Y parameter, which make the latter more difficult to test. While W can be probed in both the charged and the neutral DY channels, Y only affects the neutral DY process. Furthermore, if the neutral channel is analyzed by fitting only the invariant mass distribution, the single combination of Wilson coefficients that is probed at the linear level, in eq.~(\ref{singdir}), turns out to be proportional to $\textrm{W} + 0.6\,\textrm{Y}$ (using Table~\ref{tab:VSVD}). Therefore the sensitivity to Y is almost two times weaker than to W.

\begin{figure}
\centering
\includegraphics[width=0.635\textwidth]{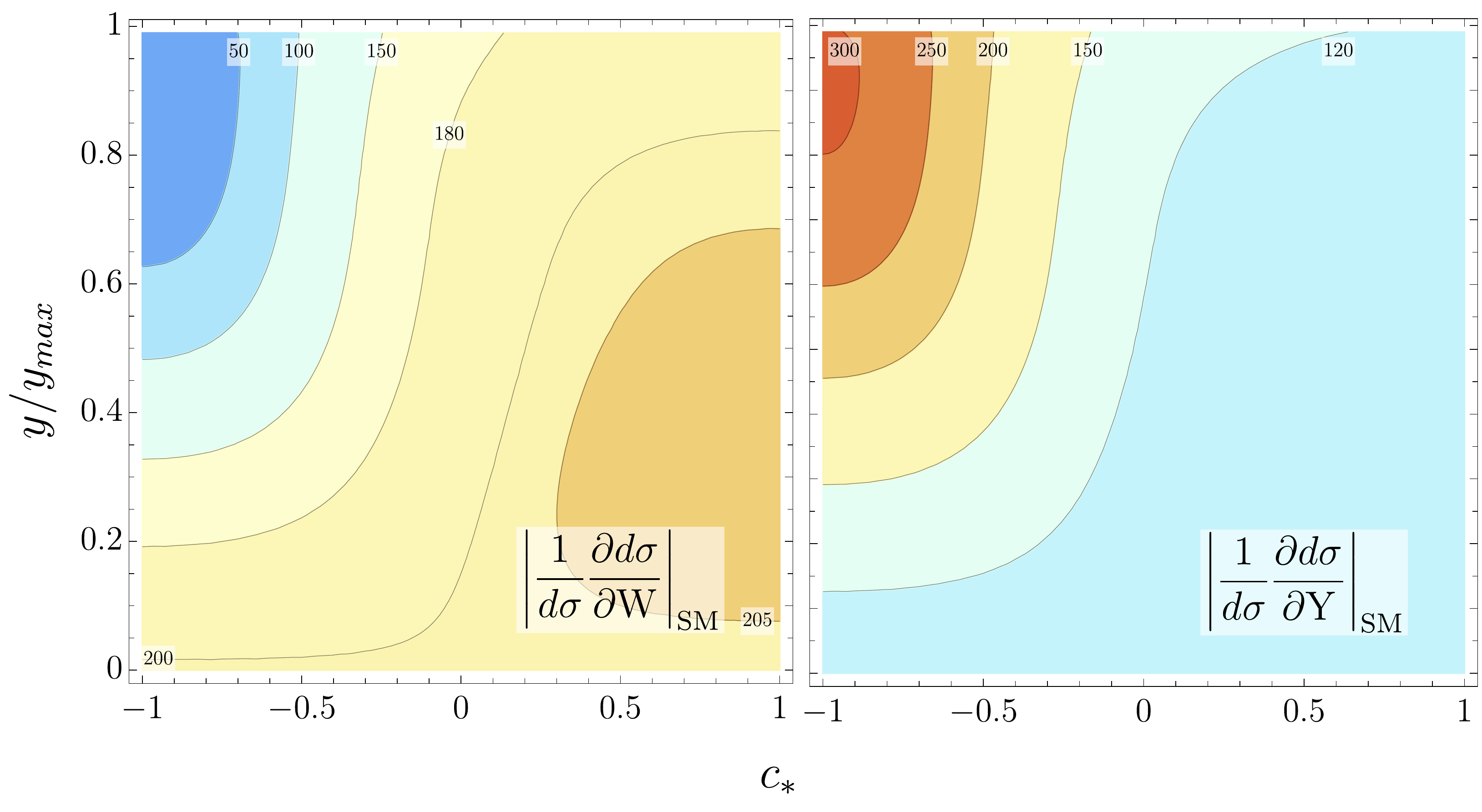}
\hfill
\includegraphics[width=0.343\linewidth]{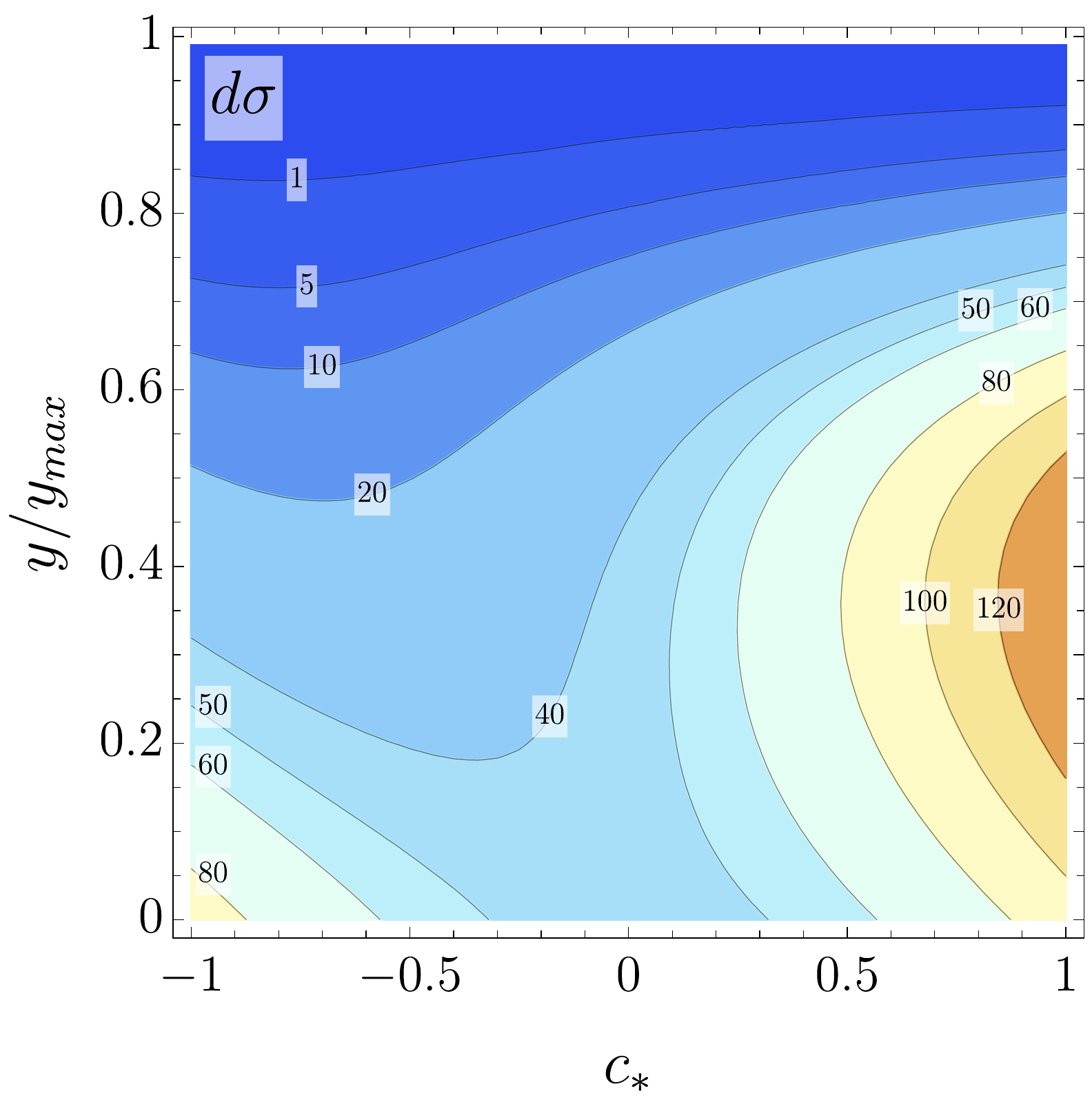}
\caption{Left panel: Absolute value of the logarithmic derivative of the tree-level differential cross-section with respect to the W and Y parameters
in the $(c_*,y)$ plane. Right panel: Level contours of the tree-level SM differential cross-section (in arbitrary units).
Both plots are obtained at fixed $m_{\ell\ell} = 1\;\textrm{TeV}$.
\label{Fig:WYlogder}}
\end{figure}

As we discussed above, a fully-differential analysis in the neutral channel can help to disentangle different new physics contributions. This happens also for the W and Y parameters. To illustrate this point we show in the left panel of Figure~\ref{Fig:WYlogder}\footnote{We employed the package of Ref.~\cite{Clark:2016jgm} to obtain semi-analytic expressions for the PDF.} the logarithmic derivatives of the tree-level differential cross-section with respect to W and Y, namely
\begin{equation}\displaystyle
    \left.\frac{1}{d \sigma} \frac{\partial d \sigma}{\partial \textrm{W}}  \right|_{\text{SM}}\,, \qquad\quad  \left.\frac{1}{d \sigma} \frac{\partial d \sigma}{\partial \textrm{Y}}  \right|_{\text{SM}}\,,
\end{equation}
evaluated at the SM point $\textrm{W}=\textrm{Y}=0$. For definiteness the dilepton invariant mass has been set to $m_{\ell\ell} = 1\;\textrm{TeV}$ in the figure. The logarithmic derivative scales like $m_{\ell\ell}^2$ as a function of the mass. As expected, in most of the kinematic space, i.e.~for small rapidity and for $c_* \gtrsim 0$, the cross-section dependence on W is roughly twice stronger than on Y. In particular this happens in the regions with larger cross-section, as can be seen from the plot in the right panel of Figure~\ref{Fig:WYlogder}.

The behavior, however, drastically changes in the corner with $c_* \lesssim -0.5$ and $y/y_{{\rm{max}}} \gtrsim 0.5$. For these configurations the differential cross-section mostly depends on Y, while the sensitivity to W is small. This feature can be easily understood from the analysis we performed in the previous section. For large rapidity and $c_* \sim -1$ the differential cross-section is controlled by $P_o^q$, which gets contributions from subprocesses with opposite fermion chiralities. Since W corresponds to an operator with only left-handed fields, it can contribute only to the same-chirality subprocesses and not to $P_o^q$. Exploiting the fully-differential distribution for the fit is thus expected to improve the determination of Y. It must be however noticed that the differential cross-section in the $y \sim y_{\rm{max}}$ and $c_* \sim -1$ corner is somewhat suppressed, and is an order of magnitude smaller than in the $c_* > 0$ region. This means that a significant improvement in the Y determination can be obtained only when a high number of signal events are collected, so that the $y \sim y_{{\rm{max}}}$ and $c_* \sim -1$ region is sufficiently populated at high $m_{\ell\ell}$. To give an idea, at the HL-LHC, out of $\sim\!12000$ SM events with $m_{\ell\ell} > 1.1\;\textrm{TeV}$, only $210$ events are expected in the region with  $y/y_{\rm{max}} > 0.4$ and $c_* < -0.6$.

\begin{figure}
\centering
\includegraphics[width=0.45\textwidth]{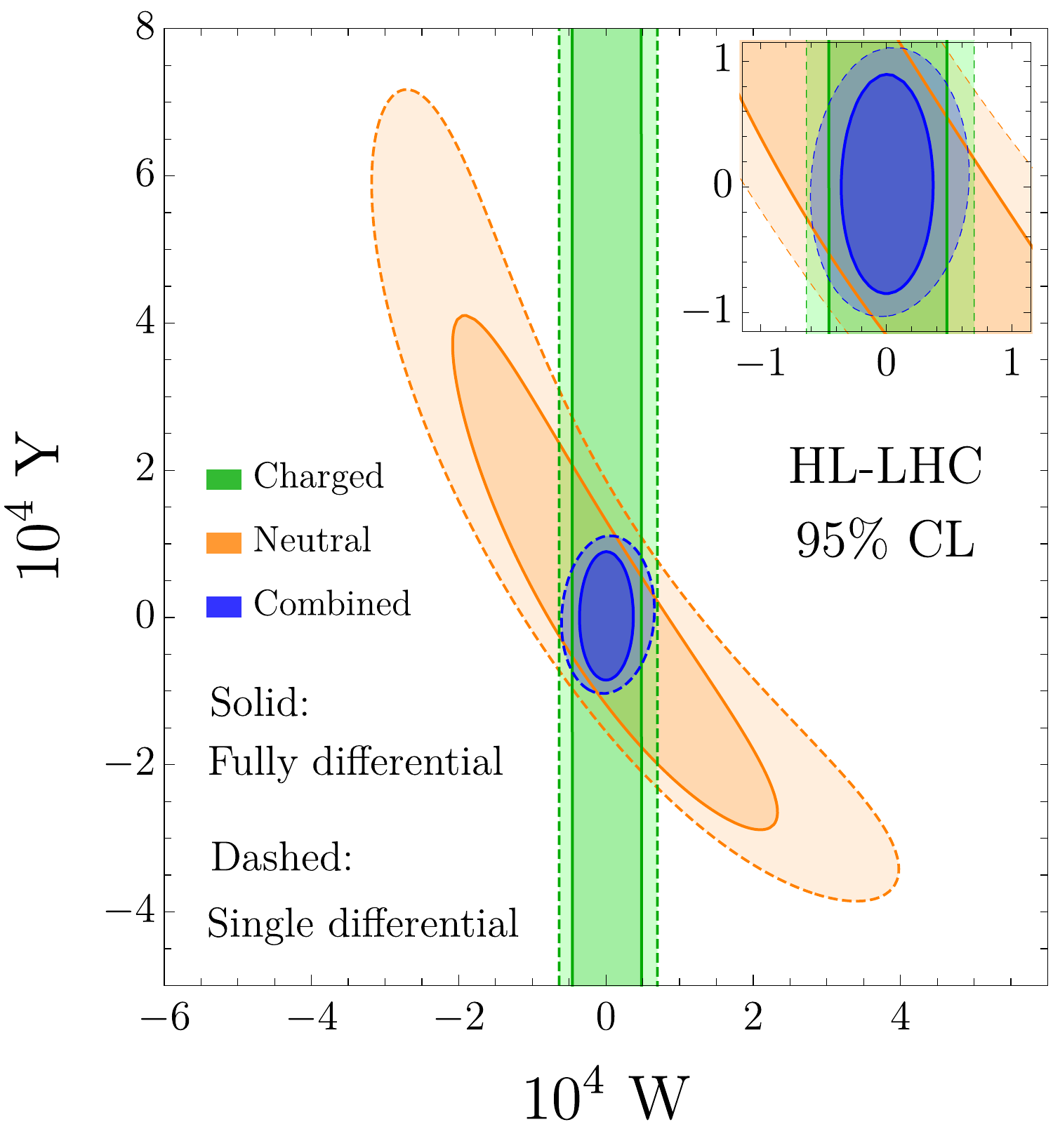}
\caption{$95{\%}$ CL allowed regions for W and Y at the $14\;\textrm{TeV}$ HL-LHC. The green and orange shaded regions correspond to the bounds from charged and neutral DY, respectively, whereas the combined bounds are given by the blue shaded regions. The fully-differential analysis results are reported with solid contours while the
dashed contours represent the sensitivity of the single-differential measurements.}\label{Fig:WYbound}
\end{figure}

We show in Figure~\ref{Fig:WYbound} the comparison of the projected exclusion reach on the W and Y parameters obtained from a fit taking into account the fully-differential distribution or the single-differential (invariant mass or transverse momentum for neutral and charged DY, respectively) distributions. To obtain the bounds we considered the HL-LHC benchmark, with collider energy $14\;\textrm{TeV}$ and ${\cal L} = 3\;\textrm{ab}^{-1}$ integrated luminosity, and we assumed that the experimental measurements of the cross-section coincide with the SM predictions.\footnote{Results for the LHC run 3 benchmark are reported in Appendix~\ref{app:LHCProj}.}
The fit of the charged DY process was obtained by considering a set of bins in the transverse momentum and rapidity of the charged lepton, whose boundaries are
\begin{eqnarray}
p_{T,\ell}: && \{150, 180, 225, 300, 400, 550, 750, 1000, 1300, 7000\}\;\textrm{GeV}\,,\\
\rule{0pt}{1.25em}\eta_l/\eta_{\rm{max}}: && \{0, 1/3, 2/3, 1\}\,,
\end{eqnarray}
where $\eta_{\rm{max}}$ is the minimum between the acceptance cut of $2.5$ and the maximal kinematically allowed rapidity $\arctanh[(1- 4 p_{T,\ell}^2/S)^{1/2}]$. For the neutral DY channel we instead used a binning in $m_{\ell\ell}$, $c_*$ and $y$, with boundaries
\begin{eqnarray}
m_{\ell\ell}: && \{300, 360, 450, 600, 800, 1100, 1500, 2000, 2600, 14000\}\;\textrm{GeV}\,,\\
\rule{0pt}{1.25em}c_*: && \{-1, -0.6, -0.2, 0.2, 0.6, 1\}\,,\\
\rule{0pt}{1.25em}y/y_{\rm{max}}: && \{0, 0.2, 0.4, 0.6, 1\}\,.
\end{eqnarray}

The cross-section predictions are obtained as in Ref.~\cite{Ricci:2020xre}, at NLO in QCD combined with parton showering (based on \PWG~\cite{Alioli:2008gx} and \PYT~\cite{Sjostrand:2014zea}) and at the NLL order in the EW expansion. The effects due to the W and Y parameters (and the EW logarithms) are included through reweighting, which enables fast and accurate Monte Carlo predictions in the relatively large number of bins (a total of $234$) that we consider in the fully-differential analysis. The $O_{2W}'$, $O_{2B}'$ operators have been defined at the renormalization scale of $10$~TeV.

The projected bounds take into account, following again Ref.~\cite{Ricci:2020xre}, the PDF uncertainties estimated through the Hessian set~\textsc{PDF4LHC15\_nlo\_30\_pdfas}~\cite{Buckley:2014ana, PDF1,PDF2,PDF3,PDF4}, and a $2\%$ luminosity uncertainty. We considered an $80\%$ reconstruction efficiency for each muon and $65\%$ for each electron. The results do not take into account any additional experimental systematic uncertainty. This is because we expect that the size and the correlation of these uncertainties will strongly depend on the binning and it will be quite different in the fully-differential measurement and in the single-differential one. Since a quantitative estimate of the uncertainties is not available, we set them to zero for a fair comparison of the two analysis procedures. A qualitative assessment of their potential impact is presented in Section~\ref{sec:uncertainty}.

From Figure~\ref{Fig:WYbound} we see that the fully-differential analysis gives a strong boost to the sensitivity of the neutral DY channel, improving in particular the
sensitivity along the $\textrm{W} + 0.6\,\textrm{Y}=0$ line that is weakly probed by the single-differential analysis as previously discussed. The charged DY sensitivity also improves. However it should be taken into account that the single-differential analysis is performed (like in Refs.~\cite{Farina:2016rws,Ricci:2020xre}) on the sum of the charge plus and charge minus cross-sections in each $p_{T,\ell}$ bin. The two charges are instead separately measured and combined in the fully-differential analysis, which is helpful to mitigate the impact of PDF uncertainties. The improvement we observe in the charged channel is partly due to this effect. 

Interestingly, the improvement of the fully-differential analysis is quite significant for the combination of the neutral and charged DY channels. The $95\%$CL single-parameter bounds from the combined fit are given by
\begin{equation}\displaystyle\label{eq:boundsWY}
\begin{array}{rll}
\textrm{W}: &\quad [-2.9, 3.0] \times 10^{-5} \quad & ([-4.9, 5.2] \times 10^{-5})\,,\\
\rule{0pt}{1.5em}\textrm{Y}: & \quad [-6.8, 7.1] \times 10^{-5} \quad & ([-8.3, 8.8] \times 10^{-5})\,,
\end{array}
\end{equation}
where the numbers in brackets correspond to the single-differential fit. The constraint on W becomes nearly a factor $2$ more stringent, whereas the determination of Y improves more mildly. It is interesting to notice that part of the improvement in the W determination does not come from the naive sum of the log likelihood for the neutral and charged processes, but is instead a consequence of the reduced impact of the PDF uncertainties. The PDF errors, in fact, are strongly correlated in the two channels, so that including both of them simultaneously in the fit allows one to distinguish their effects from the contributions due to new physics. 

\subsection{General quark-lepton interactions}\label{GCCSubSec}

We now consider the impact of the fully-differential analysis on the determination of the complete set of lepton-quark current-current operators listed in Table~\ref{tab:GenOp}. In order to make the comparison with the single-differential analysis more straightforward, it is convenient to choose a basis in the space of Wilson coefficients which is aligned with the directions that appears in the invariant mass distribution for the neutral dilepton channel. As we discussed in Section~\ref{treedits}, the $d \sigma/d m_{\ell\ell}^2$ distribution depends at the linear level only on two particular combinations of parameters, $G_s^u + G_o^u$ and $G_s^d + G_o^d$, with $G_{s,o}^{u,d}$ defined as in eq.~(\ref{P}). Moreover the ratio of up and down parton luminosities singles out one combination of Wilson coefficients~(\ref{singdir}) that is most effectively probed in the invariant mass distribution. We thus include in our basis the combination 
\begin{equation}\label{eq:GE+def}\displaystyle
G_E^+ = (G_o^u + G_s^u) + 0.5\,(G_o^d + G_s^d)\,,
\end{equation}
and the orthogonal one, which we denote by
\begin{equation}\label{eq:GE-def}\displaystyle
G_E^- = 0.5\,(G_o^u + G_s^u) - (G_o^d + G_s^d)\,.
\end{equation}
We further consider the two remaining combinations of parameters, which contribute to the fully-differential distribution at the linear level but not to the invariant mass distribution
\begin{equation}\label{eq:GOdef}\displaystyle
G_O^+ = (G_o^u - G_s^u) + 0.5\,(G_o^d - G_s^d)\,,\qquad\qquad G_O^- = 0.5\,(G_o^u - G_s^u) - (G_o^d - G_s^d)\,.
\end{equation}
Finally, we complete the seven-dimensional basis with the $\widetilde G_{lq}^{(3)}=G_{lq}^{(3)}$ coefficient and two additional combinations, $G_s^{\bot}$ and $G_o^{\bot}$, that are orthogonal to all the others. The explicit expressions are reported in Appendix~\ref{app:Form}. The $G_s^{\bot}$ and $G_o^{\bot}$ coefficients contribute (at the quadratic level) to the same-chirality and opposite-chirality subprocesses, respectively. It is important to stress that the change of basis we are performing is not orthogonal. In particular this means that the $G_{lq}^{(3)}$ coefficient of the Warsaw basis not only gives rise to $\smash{\widetilde G_{lq}^{(3)}}$ (with the same value), but also gives a correlated contribution to the $\smash{G_{E,O}^\pm}$ coefficients.

Fully-differential measurements improve the determination of the different parameters defined above, relative to the single-differential analyses, to different extents. The $\widetilde G_{lq}^{(3)}$ coefficient is mainly tested in the charged DY process, where the impact of the differential analysis is less pronounced. Its (single-operator) determination can thus improve only mildly from the combination with the neutral channel and the associated possible reduction of the impact of the PDF uncertainties. The $G_s^{\bot}$ and $G_o^{\bot}$ coefficients contribute only at quadratic order or through very small subleading terms in the $m_Z^2/m_{\ell\ell}^2$ expansion, both to the single and to the fully-differential cross-section. Therefore they will be tested with lower accuracy and they will not improve significantly with the fully-differential analysis. We thus focus on the remaining four coefficients $G_{E,O}^{\pm}$. Among those, $G_{E}^{+}$ will not improve much, since it is already effectively probed in the invariant mass distribution. A significant improvement is instead possible for the other ones.

\begin{figure}[t]
\centering
\includegraphics[width=0.7\linewidth]{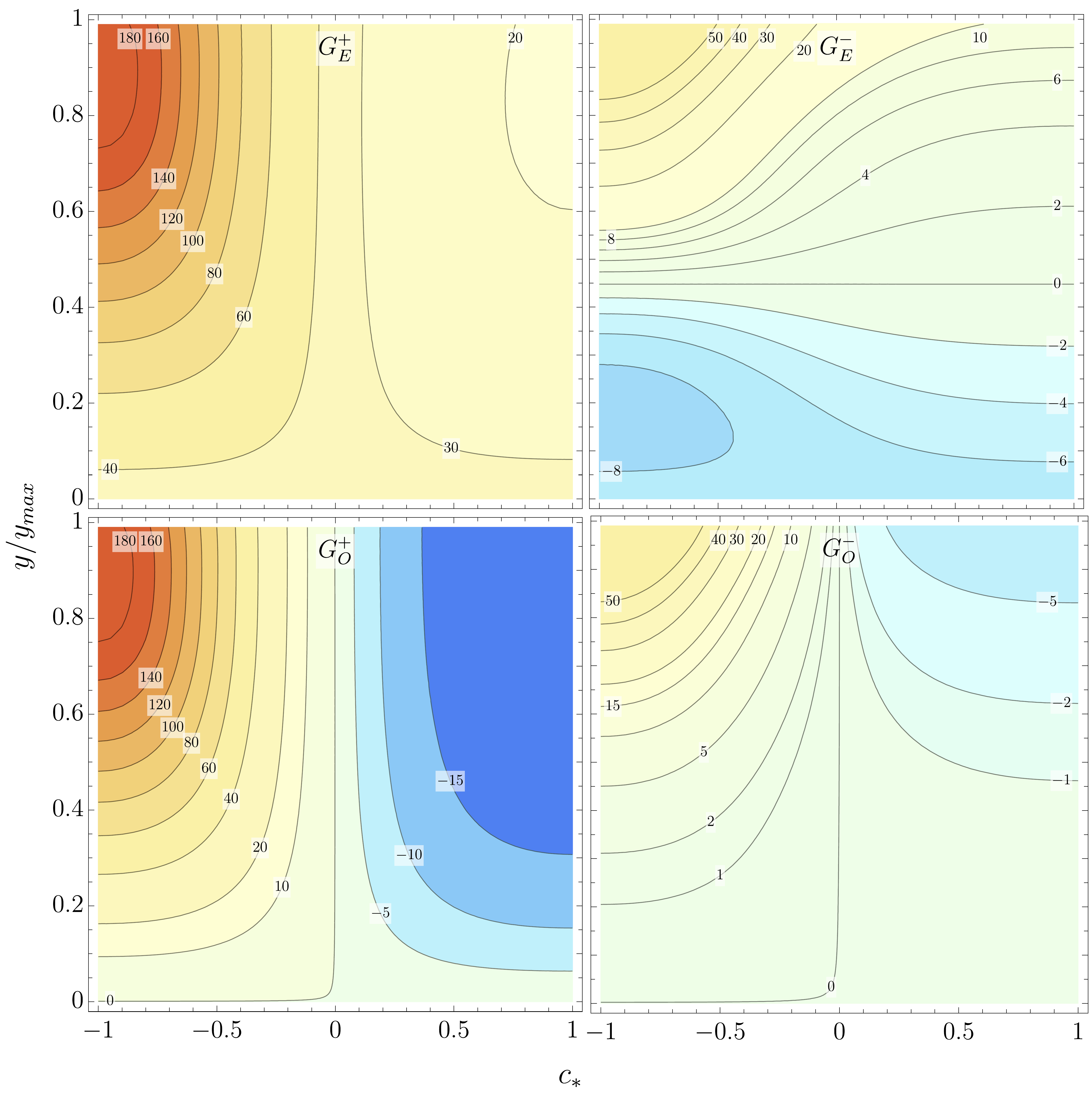}
\caption{Logarithmic derivative of the tree-level differential cross-section with respect to the $G_{E,O}^{\pm}$ coefficients
in the $(c_*,y)$ plane. The plots are obtained setting $ m_{\ell\ell}= 1$~TeV.\label{fig:log_der_4ferm}}
\end{figure}

For a first assessment of the perspectives for progress we show in Figure~\ref{fig:log_der_4ferm} the logarithmic derivative of the tree-level fully-differential cross-section for the neutral DY channel with respect to the four $G_{E,O}^{\pm}$ parameters. In the low-rapidity region ($y/y_{\rm{max}} \lesssim 0.5$) the cross-section is sensitive dominantly to $G_E^+$. This is not surprising since, as we saw in Section~\ref{treedits}, the distribution in the low-rapidity region depends on the same combination of coefficient functions that enter in the invariant-mass distribution. The high-rapidity configurations, on the contrary, show very different sensitivity patterns to the $G_{E,O}^{\pm}$ coefficients. One can see, in particular, that the $c^* > 0$ region, which has a high SM cross-section (see the right panel of Figure~\ref{Fig:WYlogder}), shows a relatively large logarithmic derivative with respect to $G_O^+$. The fully-differential analysis is therefore expected to improve significantly the determination of this coefficient and to disentangle it from $G_E^+$, which gives a different dependence of the logarithmic derivative as a function of $y$.

On the other hand, the $G_O^-$ coefficient affects the distribution mainly in the $c_* \sim -1$, $y \sim y_{\rm{max}}$ corner, in which the cross-section is rather small. For this reason we expect its determination to remain relatively poor. Finally the $G_E^-$ coefficient is in an intermediate situation. The related logarithmic derivative is significantly smaller than for $G_E^+$ and $G_O^+$, but nevertheless shows a distinctive pattern in the region with $y/y_{\rm{max}} \lesssim 0.5$, which has a good cross-section. We thus expect that the fully-differential analysis could provide some improvement on its determination.

\begin{figure}[t]
\centering
\includegraphics[width=.85\linewidth]{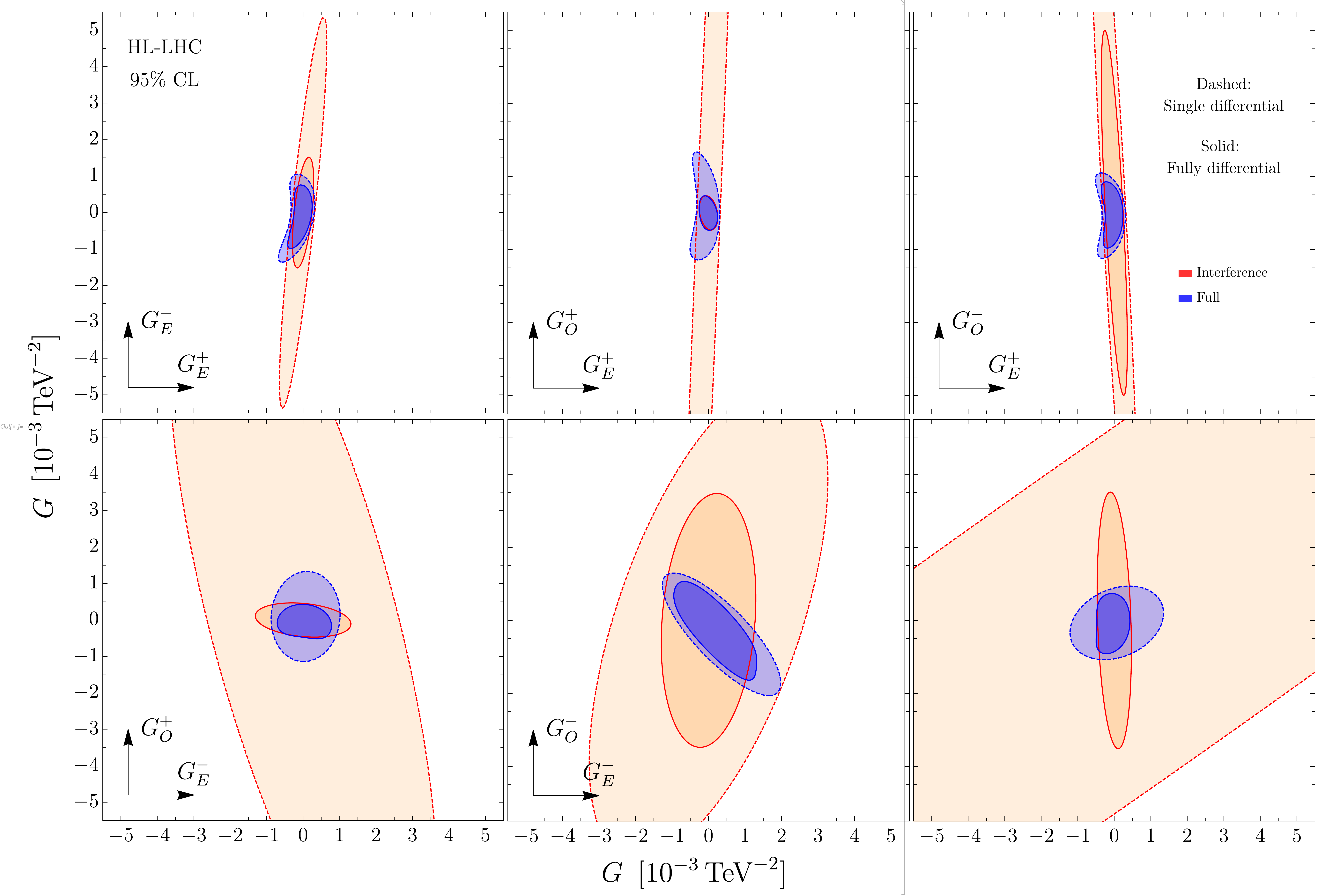}
\caption{Allowed regions at $95\%$ CL on the six coordinate planes along the four $G_{E,O}^{\pm}$ coefficients. Solid contours correspond to the fully-differential analysis, while the dashed ones are obtained with the single-differential measurements. The blue shaded regions include the full dependence on the Wilson coefficients in the cross-section, while only the linear terms are retained in the orange shaded regions.}\label{fig:6Planes}
\end{figure}

To estimate the sensitivity to $G_{E,O}^{\pm}$ we performed the same analysis presented in Section~\ref{WYSubSec} for the W and Y parameters. The two-dimensional $95\%$~CL contours for each pair of coefficients, setting the others to zero, are shown in Figure~\ref{fig:6Planes}.\footnote{The plots are obtained by combining the neutral and charged DY channels. Notice that the charged channel does not depend on the $G_E^{\pm}$ and $G_O^{\pm}$ coefficients, so its contribution is only indirect, through a reduction of the impact of the PDF uncertainties. This effect is however small.} Different sets of bounds are compared in the plots. The solid contours correspond to the $95\%$ CL constraints from the fully-differential analysis, whereas the dashed ones are obtained exploiting the invariant-mass distribution in the neutral channel and the transverse-momentum distribution (summed over the two charges as discussed in Section~\ref{WYSubSec}) in the charged channel. The blue shaded regions are obtained by considering the full dependence on the Wilson coefficients in the cross-section, while the orange shaded regions are found by taking into account only the linear terms. The axes of the ellipses for the fully-differential analysis at the linear level are aligned with the reference axes of each plane, owing to our judicious choice of the basis.

We also report, in Table~\ref{tab:SBCC}, the expected sensitivity to all the seven parameters $G_{E,O}^{\pm}$, $G_{s,o}^{\bot}$ and $\smash{\widetilde G_{lq}^{(3)}}$. We list both the single-parameter bounds obtained by setting all the others to zero and the bounds profiled over the other parameters. In the case of the fully-differential analysis we also report the results of the linearized fit.

We see from the figure and the table that the single-operator determination of $G_E^+$ is only marginally modified, with a modest improvement or order $10\%$. This was expected, as previously discussed, since the determination of $G_E^+$ from the invariant-mass distribution is already quite good. Since $G_E^+$ contributes at the linear level and it is well probed, no significant difference is present between the full fit and the linearized one in this direction.

A strong improvement is instead found in the sensitivity to $G_O^+$, as anticipated. The bound from the full fit (i.e.~including both the linear and quadratic dependence on the Wilson coefficient in the cross-section) improves roughly by a factor of $3$. The improvement in the linearized fit is even more dramatic, since $G_O^+$ does not contribute to the invariant mass distribution at the linear level up to small effects, as previously discussed. Correspondingly, an approximate flat direction is present for $G_O^+$ (see for instance the middle plot on the top row of Figure~\ref{fig:6Planes}) in the single-differential linearized contour. The fully-differential analysis is instead strongly sensitive to $G_O^+$ at the linear level and the linearized and the full fit agree very well.

The impact of the fully-differential analysis on the $G_E^-$ and $G_O^-$ parameters follows a slightly different pattern. In the full fit a mild improvement of the bounds, of order $15\%$, is found. The results, however, change drastically at the linearized level. In this case the fully-differential analysis is able to significantly improve the constraints on both parameters (see for instance the middle plot on the second row of Figure~\ref{fig:6Planes}).

\begin{table}[t]
    \centering
    {\small
    \begin{tabular}{@{\hspace{.2em}}c@{\hspace{.2em}}|c@{\hspace{.65em}}c@{\hspace{.65em}}c|c@{\hspace{.65em}}c@{\hspace{.65em}}c@{\hspace{.2em}}}
    {\bf{95\%CL}}&\multicolumn{3}{c|}{single parameter}&\multicolumn{3}{c}{profiled}\\
    \rule[-.6em]{0pt}{1.85em}$[10^{-3}\,\textrm{TeV}^{-2}]$ & fully diff. & fully diff. lin. & {\rm{single diff.}} & fully diff. & fully diff. lin. & single diff.\\
      \hline
    \rule{0pt}{1.35em}$\widetilde{G}_{lq}^{(3)}$ & $[-0.36 ,\, 0.35]$ & $[-0.36,\, 0.36]$ & $[-0.49 ,\, 0.50]$
    & $[-0.53 ,\, 0.48]$ & $[-0.57,\, 0.57]$ & $[-0.97 ,\, 0.77]$\\
    \rule{0pt}{1.35em}${G}_{E}^{+}$ & $[ -0.20,\, 0.19]$ & $[-0.20,\, 0.20]$ &  $[-0.27 ,\, 0.25]$
        & $[-0.55 ,\, 0.32]$ & $[-0.60,\, 0.60]$ & $[-1.19 ,\, 0.45]$\\
    \rule{0pt}{1.35em}${G}_{E}^{-}$ & $[-0.61 ,\, 0.65]$ & $[-1.02,\, 1.02]$ & $[-0.77 ,\, 0.90]$
        & $[-1.52 ,\, 1.30]$ & $[-1.62,\, 1.62]$ & $[-3.31 ,\, 1.94]$\\
    \rule{0pt}{1.35em}${G}_{O}^{+}$ &  $[-0.38,\, 0.35]$ & $[-0.37,\, 0.37]$ & $[-1.01 ,\, 1.19]$
        & $[-0.44 ,\, 0.60]$ & $[-0.82,\, 0.82]$ & $[-1.32,\, 2.17]$\\
    \rule{0pt}{1.35em}${G}_{O}^{-}$ & $[-0.77,\, 0.65]$ & $[-2.74,\, 2.74]$ & $[-0.95 ,\, 0.79]$
        & $[-1.58,\, 1.80]$ & $[-5.85,\, 5.85]$ & $[-2.06 ,\, 3.20]$\\
    \rule{0pt}{1.35em}${G}_s^{\bot}$ & $[-2.14,\, 1.44]$ & $[-3.74,\, 3.74]$ & $[-2.24 ,\, 1.59]$
        & $[-2.46 ,\, 2.19]$ & $[-10.5,\, 10.5]$ & $[-2.91,\, 2.41]$\\
     \rule[-.75em]{0pt}{2.1em}${G}_o^{\bot}$ & $[-0.69,\, 0.67]$ & $[-18.6,\, 18.6]$ & $[-0.85 ,\, 0.84]$
        & $[-0.98,\, 0.89]$ & $[-25.5,\, 25.5]$ & $[-1.26,\, 1.31]$
    \end{tabular}
   }
       \caption{$95{\%}$ CL projected bounds (in $10^{-3}\;\textrm{TeV}^{-2}$ units) for the seven coefficients $\widetilde{G}_{lq}^{(3)}$, $G_{E,O}^\pm$ and $G_{s,o}^\bot$. The first three bounds correspond to single-operator fits, in which all other parameters are set to zero, while the last three are profiled over the other parameters. For each set of bounds the three columns correspond to the complete fully-differential fit, the linearized one and the single-differential measurement fit.}
        \label{tab:SBCC}
    \end{table}

The profiled bounds reported in Table~\ref{tab:SBCC} are more difficult to interpret. They significantly differ from the single operator ones, signaling the presence of non-negligible correlations among the various parameters. We notice that for many parameters the fully-differential analysis improves the profiled bound more than the single-operator one. This pattern is particularly visible for the $\widetilde G_{lq}^{(3)}$, $G_E^+$ and $G_E^-$ parameters, and, to a lesser degree, for $G_O^+$ and $G_O^-$. The origin of this behavior can be traced back to the reduction of flat directions in the fully-differential fit, which helps in reducing the correlations among the various Wilson coefficients.

\begin{table}[t]
    \centering
    {\small
    \begin{tabular}{@{\hspace{.2em}}c@{\hspace{.2em}}|c@{\hspace{.65em}}c@{\hspace{.65em}}c|c@{\hspace{.65em}}c@{\hspace{.65em}}c@{\hspace{.2em}}}
     {\bf{95\%CL}}&\multicolumn{3}{c|}{single parameter}&\multicolumn{3}{c}{profiled}\\
    \rule[-.6em]{0pt}{1.85em}$[10^{-3}\,\textrm{TeV}^{-2}]$ & fully diff. & fully diff. lin. & single diff. & fully diff. & fully diff. lin. & single diff.\\
    \hline
    \rule{0pt}{1.35em}${G}_{lq}^{(3)}$ & $[-0.46,\, 0.44]$ & $[-0.45,\, 0.45]$ & $[-0.80,\, 0.75]$
    & $[-0.53,\, 0.48]$ & $[-0.57,\, 0.57]$ & $[-0.97,\, 0.77]$\\
    \rule{0pt}{1.35em}${G}_{lq}^{(1)}$ & $[-1.95,\, 2.42]$ & $[-2.15,\, 2.15]$ & $[-2.42,\, 3.55]$
        & $[-4.69,\, 5.28]$ & $[-12.8,\, 12.8]$ & $[-5.93,\, 8.75]$\\
    \rule{0pt}{1.35em}${G}_{qe}$ & $[-2.13,\, 3.29]$ & $[-2.56,\, 2.56]$ & $[-3.19,\, 6.19]$
        & $[-4.31,\, 5.35]$ & $[-139,\, 139]$ & $[-7.11,\, 7.05]$\\
    \rule{0pt}{1.35em}${G}_{lu}$ & $[-1.76,\, 2.12]$ & $[-1.92,\, 1.92]$ & $[-2.87,\, 4.22]$
        & $[-3.18,\, 5.41]$ & $[-69.2,\, 69.2]$ & $[-7.38,\, 8.57]$\\
    \rule{0pt}{1.35em}${G}_{ld}$ & $[-8.98,\, 5.01]$ & $[-7.09,\, 7.09]$ & $[-11.9,\, 6.92]$
        & $[-9.89,\, 8.95]$ & $[-148,\, 148]$ & $[-12.6,\, 14.0]$\\
    \rule{0pt}{1.35em}${G}_{eu}$ & $[-1.22,\, 1.30]$ & $[-1.26,\, 1.26]$ & $[-1.51,\, 1.65]$
        & $[-3.38,\, 5.75]$ & $[-16.3,\, 16.3]$ & $[-4.46,\, 7.29]$\\
    \rule[-.75em]{0pt}{2.1em}${G}_{ed}$ & $[-4.74,\, 3.55]$ & $[-4.03,\, 4.03]$ & $[-7.12,\, 4.47]$
        & $[-9.25,\, 9.33]$ & $[-41.3,\, 41.3]$ & $[-10.7,\, 12.7]$
    \end{tabular}}
     \caption{$95{\%}$ CL projected bounds for the four-fermion operator coefficients in the Warsaw basis. The bounds are given in $10^{-3}\;\textrm{TeV}^{-2}$ units.
        }
        \label{tab:Bounds_Warsaw}
\end{table}    

For completeness, we report in Table~\ref{tab:Bounds_Warsaw} the bounds on the four-fermion operators in the Warsaw basis. In this basis we find that $G_{lq}^{(3)}$ is expected to be determined with much higher precision than the other parameters. Moreover its determination is only mildly affected by profiling, differently from the bounds on the other coefficients that significantly degrade in the profiled fit. This behavior is clearly due to the fact that $G_{lq}^{(3)}$ is tested with high precision in the charged DY channel, which is not affected by the other effective operators. The impact of a fully-differential analysis is quite large for many Warsaw operators. In particular the bounds on $G_{lq}^{(3)}$, $G_{qe}$ and $G_{lu}$ become roughly $40\%$ tighter, while the constraints on the other operators improve by an amount of order $10 - 20\%$.

\begin{figure}[t]
\centering
\includegraphics[width=.52\linewidth]{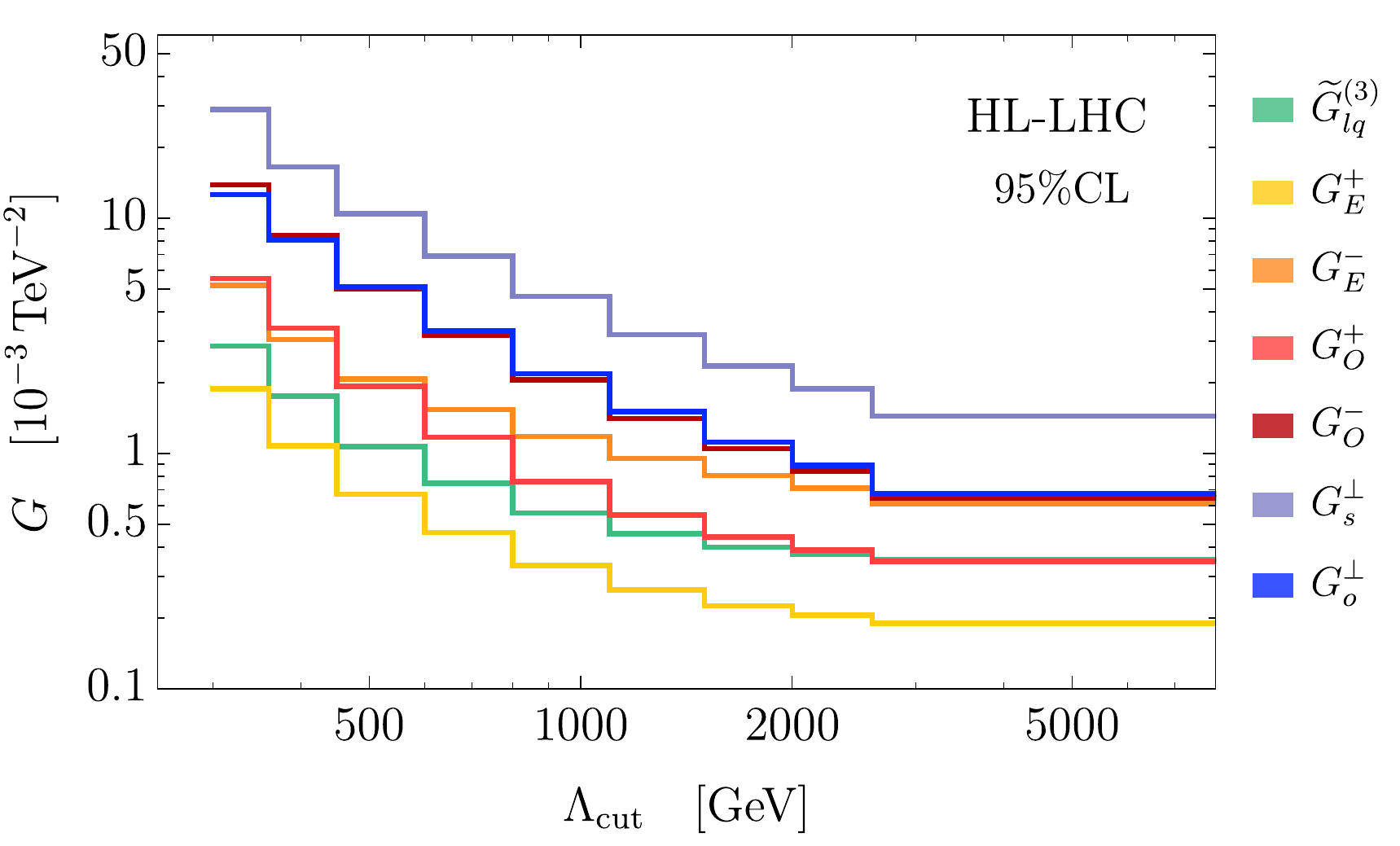}
\hfill
\raisebox{1.2em}{\includegraphics[width=.46\linewidth]{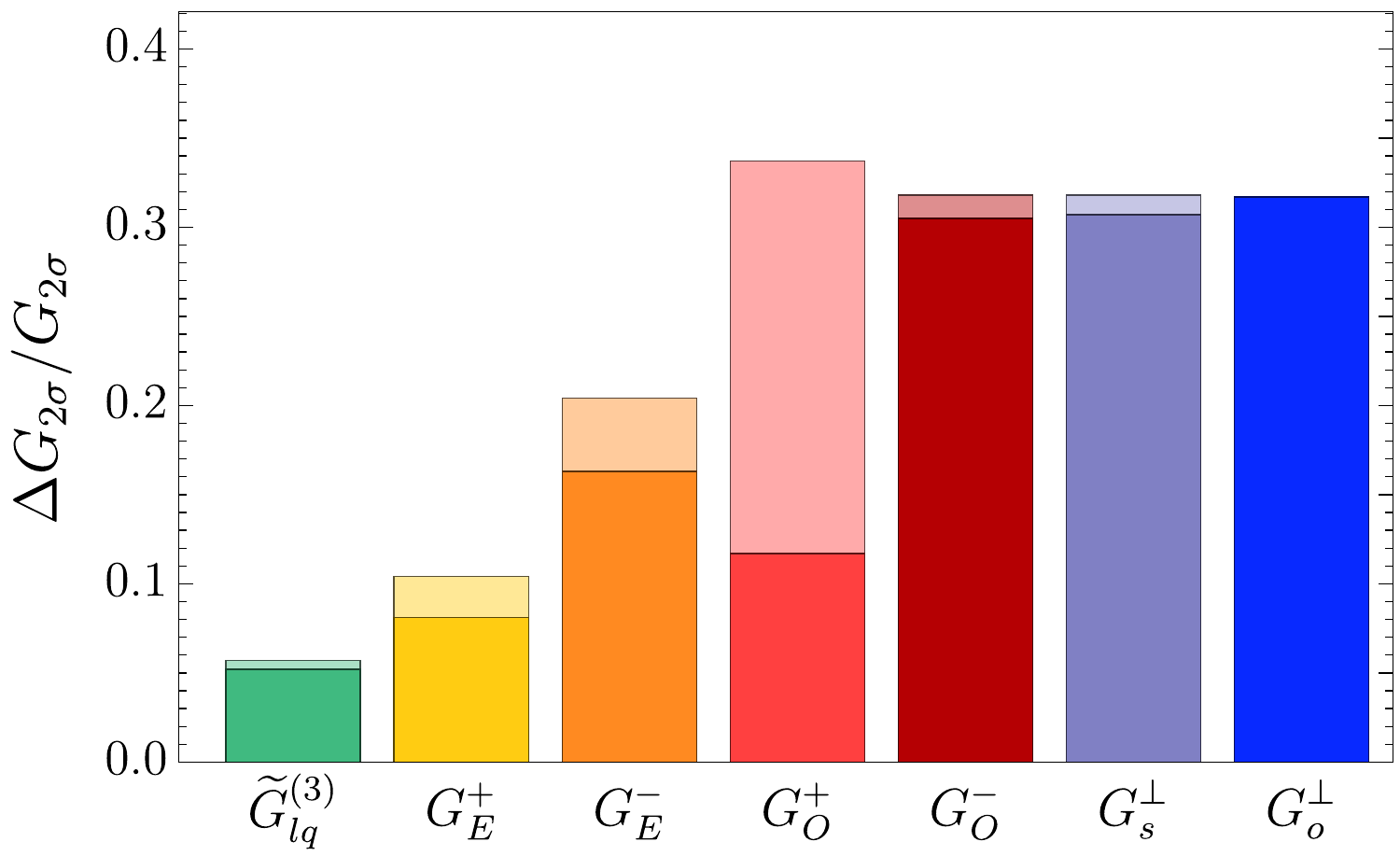}}
\caption{Left panel: Expected bounds on the Wilson coefficients as a function of the upper cut on the energy of the events $\Lambda_{\rm{cut}}$. Right panel: Relative change in the bounds in the last two bins in $\Lambda_{\rm{cut}}$. The darker shadowing corresponds to the fully-differential analysis, while the lighter shadowing corresponds to the {\rm{single diff.}} analysis.}\label{fig:Ladder}
\end{figure}

As a last point, we investigate the dependence of our results on the maximal energy scale, $\Lambda_{\rm{cut}}$, of the measurements included in the fit. This gives useful indications on the measurements that contribute more to the final sensitivity and on the energy range of validity of the EFT description of new physics that is theoretically required for the bounds to apply. Following Refs.~\cite{Farina:2016rws,Ricci:2020xre}, we show on the left panel of Figure~\ref{fig:Ladder} how the single-operator bounds change by retaining in the fit only the bins where $m_{\ell\ell} < \Lambda_{\rm{cut}}$ in the neutral channel and $p_{T,\ell} < \Lambda_{\rm{cut}}/2$ in the charged one. We see that the bounds on the Wilson coefficients with more stringent constraints, namely $\widetilde G_{lq}^{(3)}$, $G_E^+$ and $G_O^+$, saturate around $\Lambda_{\rm{cut}} \sim 2\;\textrm{TeV}$. In particular, removing the last bin (starting at $2600$~GeV) has an extremely mild impact. The constraints on the other coefficients, on the contrary, receive sizable contributions from the events in the last bin. This behavior can be explained by recalling that the bounds on the $\widetilde G_{lq}^{(3)}$, $G_E^+$ and $G_O^+$ coefficients in the fully-differential analysis are mainly driven by the linear interference terms in the cross-section prediction. These terms grow linearly with the partonic $\hat s$ so that the energy region with good sensitivity, which we find to be $\sqrt{\hat s} \sim 1 - 2\;\textrm{TeV}$, is where this growth starts being balanced by the decrease of the quark luminosity. For the other parameters, instead, the bounds are driven mostly by the square of the BSM contributions, which grow like $\hat s^2$. The faster growth pushes the sensitive region to higher energies.

To appreciate better this point, we show in the right panel of Figure~\ref{fig:Ladder} the relative change in the bounds on the various parameters when the last bin is removed. For the fully-differential fit (darker shadowing), a variation below around $10\%$ is observed for $\smash{\widetilde G_{lq}^{(3)}}$, $G_E^+$ and $G_O^+$. The $\smash{G_O^-}$, $\smash{G_s^\bot}$ and $\smash{G_o^\bot}$ coefficients, instead, show a relatively large change in the bounds ($\simeq 30\%$). Finally, the $G_E^-$ coefficient shows an intermediate behavior, which is explained by the fact that for its determination the linear interference terms and the quadratic terms have comparable weight. It is interesting to notice that the sensitivity to the last bin of $G_O^+$ is quite lower in the fully-differential fit than in the single-differential one (displayed with lighter shadowing in the figure). This is because in the former case the bound is driven by the linear terms, while in the latter it is mainly driven by the quadratic terms. This difference can be also seen in the single-parameter bounds on $G_O^+$, which improve by roughly a factor $3$ with the fully-differential analysis (see Figure~\ref{fig:6Planes} and Table~\ref{tab:SBCC}).

\subsubsection*{Running Effects}

Our cross-section predictions include EW corrections at the single-log accuracy, among which the ones associated with the Renormalization Group evolution of the EFT operators~\cite{Ricci:2020xre}.\footnote{The tools in Ref.~\cite{Celis:2017hod} have been employed for the implementation.} Therefore our results depend, in line of principle, on the operator renormalization scale. This has been set to $2$~TeV because the measurements at that scale dominate the sensitivity as previously shown. However the running effects are extremely small and our results do not depend on this choice in practice. This has been verified by repeating the fit in two ways. In one case we switched off completely the running, while in the other one we fixed the values of the Wilson coefficients at an energy scale $E = 10\;\textrm{TeV}$. In both cases the bounds on the Wilson coefficients, both single-operator and the profiled ones, change at most by $\textit{few}\;\%$. Our results can thus be safely applied even to EFT operators defined at several tens of TeV.

It is important to keep in mind that only quark-lepton current-current operators are included in our calculation. Other EFT operators do not produce growing-with-energy effects in high-energy DY, therefore their contribution is very suppressed relative to the quark-lepton ones and completely negligible if their size is not anomalously large. In particular this means that operators induced by the quark-lepton ones through running are completely negligible because running is itself a small effect. The contribution to the running of the quark-lepton operators by the other ones is also negligible, for the same reason. On the other hand, one cannot firmly exclude the presence of other EFT operators with anomalously large coefficients that are not already excluded or that can not be probed with other LHC measurements. Such operators, if found to exist after a more systematic global exploration of the LHC EFT potential, should be included in the predictions.

\subsection{Experimental uncertainties}\label{sec:uncertainty}

This section is devoted to uncertainties that are not included in the analysis presented above, namely the presence of systematic uncertainties in the experimental cross-section measurements other than the luminosity uncertainty that was already taken into account. 

Like in Ref.~\cite{Ricci:2020xre}, our fits are based on the Poisson likelihood, which takes automatically into account the statistical component of the cross-section measurement errors. The systematic component of the experimental error is instead incorporated by nuisance parameters on the expected Poisson countings. Only the nuisance corresponding to a $2\%$ luminosity uncertainty has been included in the analysis and its effect is very small as expected. The dominant experimental errors are indeed those that, unlike the luminosity, distort the shape of the differential distributions~\cite{Ricci:2020xre}. We cannot rely on any estimate of the size of these uncertainties, nor of their correlations across different analysis bins which on the other hand are expected to have a major impact on the sensitivity to new physics. In order to get a feeling of their possible impact, we adopt a crude parametrization of these effects by introducing a fully-uncorrelated $2\%$ error in all bins, both for the fully-differential and for the single-differential fit.

Including these uncorrelated systematic errors, the combined $95\%$~CL bound on the W and Y parameters become 
\begin{equation}\displaystyle
\begin{array}{rll}
{\rm{W}}: &\quad [-3.6, 3.7] \times 10^{-5} \quad & ([-5.3, 5.8] \times 10^{-5})\\
\rule{0pt}{1.5em}{\rm{Y}}: & \quad [-8.4, 8.9] \times 10^{-5} \quad & ([-10.9, 12.0] \times 10^{-5})
\end{array}
\qquad \textrm{with}\ 2\%\ \textrm{uncorr. syst.}\,.
\end{equation}
The numbers in parentheses refer to the single-differential fit. Comparing with the results in eq.~(\ref{eq:boundsWY}), we see that the bounds from the fully-differential analysis become roughly $25\%$ weaker. In the case of the single-differential analysis, the bounds on Y suffer from a similar change, while the ones on W are less affected and are only $10\%$ weaker. The advantage of a fully-differential analysis is however still evident also in these results.

\begin{table}[t]
    \centering
    {\small
    \begin{tabular}{@{\hspace{.2em}}c@{\hspace{.2em}}|c@{\hspace{.65em}}c@{\hspace{.65em}}c|c@{\hspace{.65em}}c@{\hspace{.65em}}c@{\hspace{.2em}}}
     {\bf{95\%CL}}&\multicolumn{3}{c|}{single parameter}&\multicolumn{3}{c}{profiled}\\
    \rule[-.6em]{0pt}{1.85em}$[10^{-3}\,\textrm{TeV}^{-2}]$ & fully diff. & fully diff. lin. & single diff. & fully diff. & fully diff. lin. & single diff.\\
     \hline
    \rule{0pt}{1.35em}$\widetilde{G}_{lq}^{(3)}$ & $[-0.46, 0.46]$&$ [-0.46, 0.46]$&$[-0.62, 0.63]$&$[-0.68, 0.59]$&$[-0.72,0.72]$&$[-1.06, 0.84]$\\
    \rule{0pt}{1.35em}${G}_{E}^{+}$ & $[-0.25, 0.23]$&$[-0.24, 0.24]$&$[-0.38, 0.32]$&$[-0.67, 0.36]$&$[-0.73, 0.73]$&$[-1.51, 0.51]$\\
    \rule{0pt}{1.35em}${G}_{E}^{-}$ & $[-0.67, 0.70]$&$[-1.45, 1.45]$& $[-0.85, 0.95]$&$[-1.80, 1.61]$&$[-1.96, 1.96]$&$[-3.73, 2.24]$\\
    \rule{0pt}{1.35em}${G}_{O}^{+}$ &  $[-0.41, 0.37]$&$[-0.40, 0.40]$&$[-1.07, 1.24]$&$[-0.48, 0.72]$&$[-0.97,0.97]$&$[-1.43, 2.49]$\\
    \rule{0pt}{1.35em}${G}_{O}^{-}$ & $[-0.79, 0.67]$&$[-3.01, 3.01]$&$[-0.99, 0.83]$&$[-1.84, 2.00]$&$[-6.21, 6.21]$&$[-2.33, 3.61]$\\
    \rule{0pt}{1.35em}${G}_s^{\bot}$ & $[-2.18, 1.50]$&$[-4.21, 4.21]$&$[-2.31, 1.70]$&$[-2.59, 2.44]$&$[-11.9,11.9]$&$[-3.17, 2.85]$\\
     \rule[-.75em]{0pt}{2.1em}${G}_o^{\bot}$ & $[-0.71, 0.69]$&$[-22.2, 22.2]$&$[-0.89, 0.88]$&$[-1.03, 0.91]$&$
[-31.8, 31.8]$&$ [-1.47, 1.46]$
    \end{tabular}
    }
    \caption{$95{\%}$ CL projected bounds for the seven coefficients $\widetilde{G}_{lq}^{(3)}$, $G_E^\pm$, $G_O^\pm$ and $G_{s,o}^\bot$, including a $2\%$ uncorrelated systematic uncertainty.}
    \label{tab:SOR}
\end{table}

\begin{table}[t]
    \centering
    {\small
    \begin{tabular}{@{\hspace{.2em}}c@{\hspace{.2em}}|c@{\hspace{.65em}}c@{\hspace{.65em}}c|c@{\hspace{.65em}}c@{\hspace{.65em}}c@{\hspace{.2em}}}
     {\bf{95\%CL}}&\multicolumn{3}{c|}{single parameter}&\multicolumn{3}{c}{profiled}\\
    \rule[-.6em]{0pt}{1.85em}$[10^{-3}\,\textrm{TeV}^{-2}]$ & fully diff. & fully diff. lin. & single diff. & fully diff. & fully diff. lin. & single diff.\\
    \hline
    \rule{0pt}{1.35em}${G}_{lq}^{(3)}$&$[-0.57, 0.55]$&$ [-0.56, 0.56]$&$[-0.89, 0.82]$&$ [-0.68, 0.59]$&$ [-0.72, 0.72]$&$[-1.06, 0.84]$\\
    \rule{0pt}{1.35em}${G}_{lq}^{(1)}$ & $[-2.21, 2.97]$&$[-2.50, 2.50]$&$[-2.90, 6.60]$&$ [-5.28, 5.88]$&$[-14.9,14.9]$&$[-6.65, 9.92]$\\
    \rule{0pt}{1.35em}${G}_{qe}$ &$[-2.39, 4.33]$&$[-3.02, 3.02]$&$[-3.78, 7.83]$&$[-4.40, 5.67]$&$[-174,174]$&$ [-7.70, 8.19]$\\
    \rule{0pt}{1.35em}${G}_{lu}$ & $[-1.98, 2.53]$&$[-2.21, 2.21]$&$[-3.49, 8.10]$&$[-3.29, 5.74]$&$[-86.7,86.7]$&$[-7.95, 9.81]$\\
    \rule{0pt}{1.35em}${G}_{ld}$ & $[-10.1, 5.58]$&$[-8.83, 8.83]$&$[-13.1, 8.00]$&$ [-10.57, 9.19]$&$[-179,179]$&$ [-15.0, 15.3]$\\
    \rule{0pt}{1.35em}${G}_{eu}$ & $[-1.48, 1.63]$&$[-1.55, 1.55]$&$[-1.95, 2.30]$&$[-3.46, 6.58]$&$[-18.7,18.7]$&$[-4.65, 8.67]$\\
    \rule[-.75em]{0pt}{2.1em}${G}_{ed}$ & $[-7.83, 4.39]$&$[-5.42, 5.42]$&$[-12.7, 5.62]$&$[-11.0, 9.81]$&$[-46.5, 46.5]$&$ [-13.3, 13.5]$
    \end{tabular}
    }
    \caption{$95{\%}$ CL projected bounds for the four-fermion operator coefficients in the Warsaw basis, including a $2\%$ uncorrelated systematic uncertainty.}
\label{tab:SWB}
\end{table}    

In Tables~\ref{tab:SOR} and~\ref{tab:SWB} we give the bounds on the lepton-quark four-fermion operators including the uncorrelated systematic uncertainty. The impact of the uncertainty is relatively large on $\widetilde G_{lq}^{(3)}$ and $G_E^+$, whose determination becomes roughly $25\%$ weaker. The reduction in sensitivity on the other coefficients is instead milder, at most of order $10\%$. Similar results are found for the operators in the Warsaw basis (Table~\ref{tab:SWB}). In this basis the most affected operators are $G_{lq}^{(3)}$, $G_{eu}$ and $G_{ed}$, with a loss of sensitivity of order $25\%$, while the bounds on the other operators are quite stable. One can also see that the impact of the systematic uncertainty on the fully-differential fit and on the single-differential one is comparable.

\section{Sensitivity to minimal Z$^\mathbf{\prime}$ models}\label{sec:Zprime}

For a concrete assessment of the benefits of the fully-differential analysis, we consider in this section a minimal BSM scenario featuring a single additional vector boson that gauges a generic linear combination of the hypercharge $\textrm{U}(1)_Y$ and $B-L$. The Lagrangian describing the new vector boson is
\begin{align}
\mathcal{L}_{Z'} = -\frac{1}{4}F_{\mu \nu}^2+ \frac{1}{2} {M}^2 A^2 + A_{\mu} J^{\mu}\,,
\end{align}
with the current
\begin{equation}\displaystyle
J^{\mu} = \sum_{f}\left[g_Y Y(f)+g_{BL}(B-L)(f)\right]\bar{f} \gamma^{\mu} f
+ g_Y\frac{i}{2} H^{\dagger} {\scriptstyle \overleftrightarrow{\rule{0pt}{.75em}}}\hspace{-.95em}{D} H\,,
\end{equation}
where $f$ denote the SM fermions and $H$ is the Higgs doublet. In the above formula $g_Y$ and $g_{BL}$ are free parameters, while $Y(f)$ and $B(f)$ and $L(f)$ are the hypercharge, the baryon and the lepton numbers of the various fermions, respectively. This model has been studied extensively in the literature and in particular in Ref.~\cite{Salvioni:2009mt}, where a first projection of LHC direct searches sensitivity was given, and compared with the indirect constraints from precision measurements (EWPT) performed at LEP and other experiments.\footnote{For the model to be free of gauge anomalies, three right handed neutrinos with $B-L$ coupling must be present. Contrary to Ref.~\cite{Salvioni:2009mt}, we take these states to be nearly massless.}

\begin{figure}[t]
\centering
\includegraphics[width=.515\linewidth]{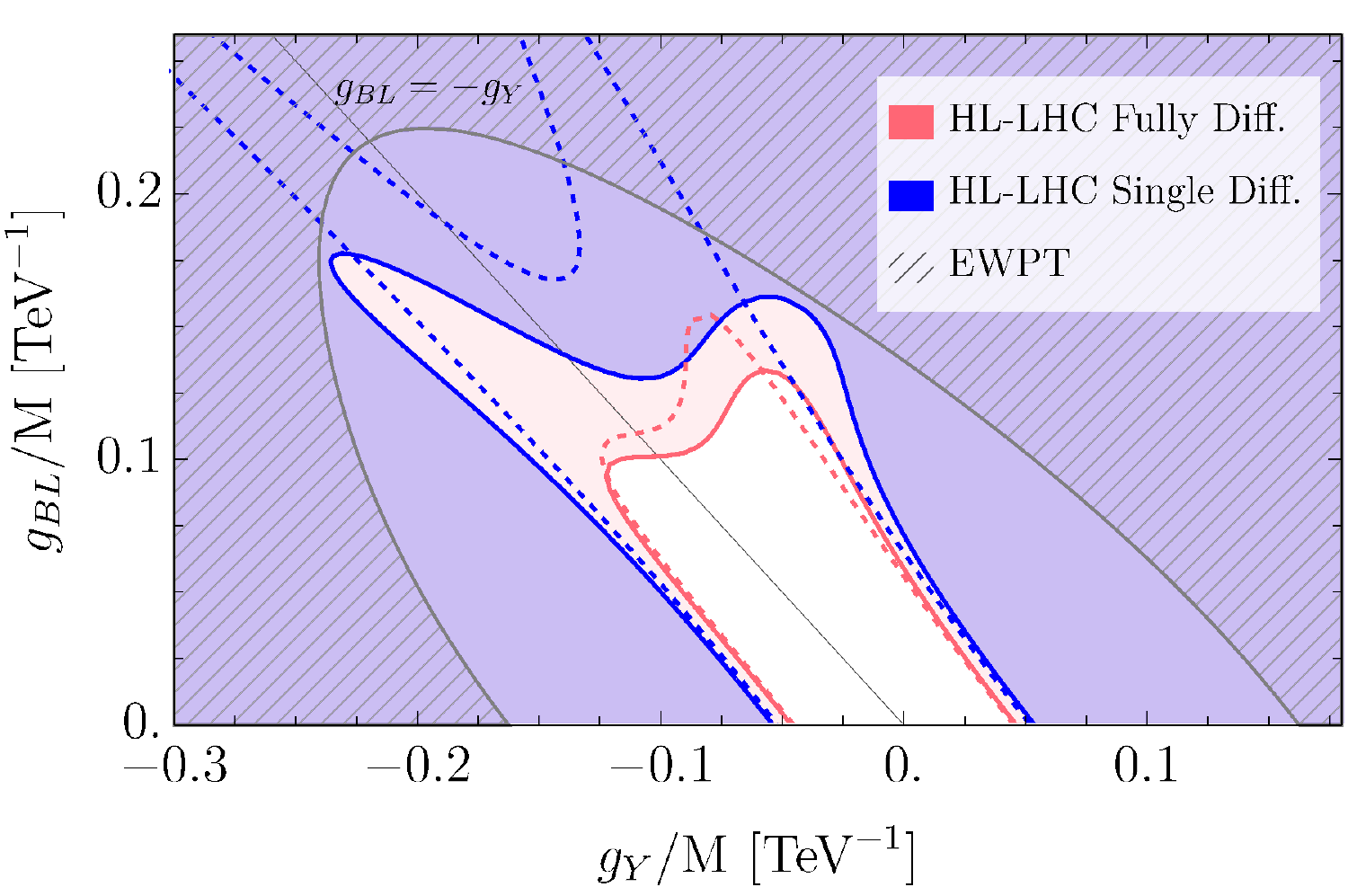}
\hfill
\includegraphics[width=.465\linewidth]{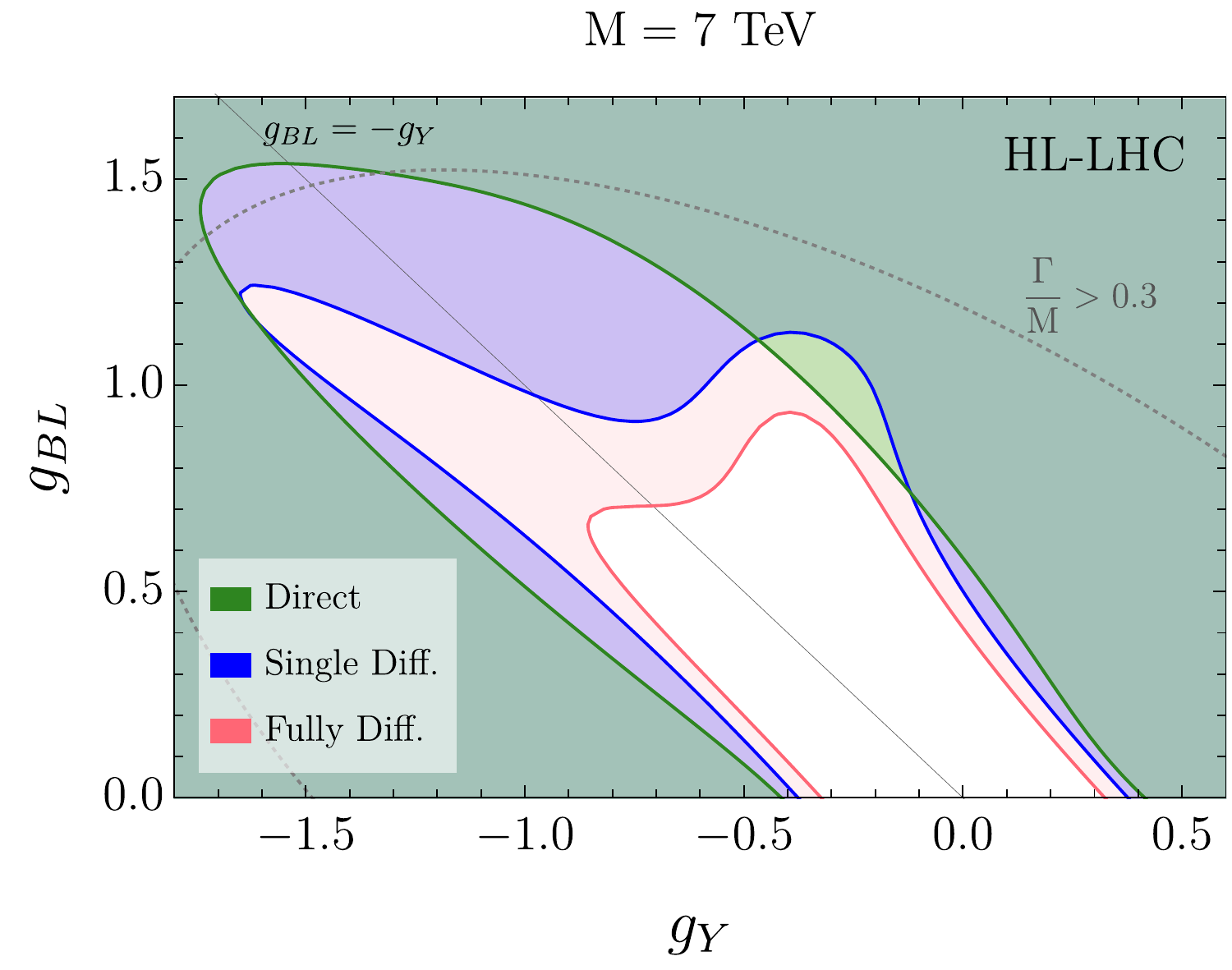}
\caption{Left panel: $95\%$~CL exclusion reach in the $(g_{Y}/M, g_{BL}/M)$ plane from EWPT (gray shaded region)~\cite{Salvioni:2009mt} and projected exclusion at the HL-LHC (blue and orange shaded regions). Right panel: Comparison of direct (green shaded region) and indirect (blue and orange shaded regions) exclusions at the HL-LHC for a heavy vector of mass $M = 7\;\textrm{TeV}$. In both panels the blue shaded region corresponds to the fully-differential fit, while the orange shaded region is obtained with the single-differential one.}
\label{fig:Zprime_comp}
\end{figure}

\begin{figure}[t]
    \centering
    \includegraphics[width=0.32\linewidth]{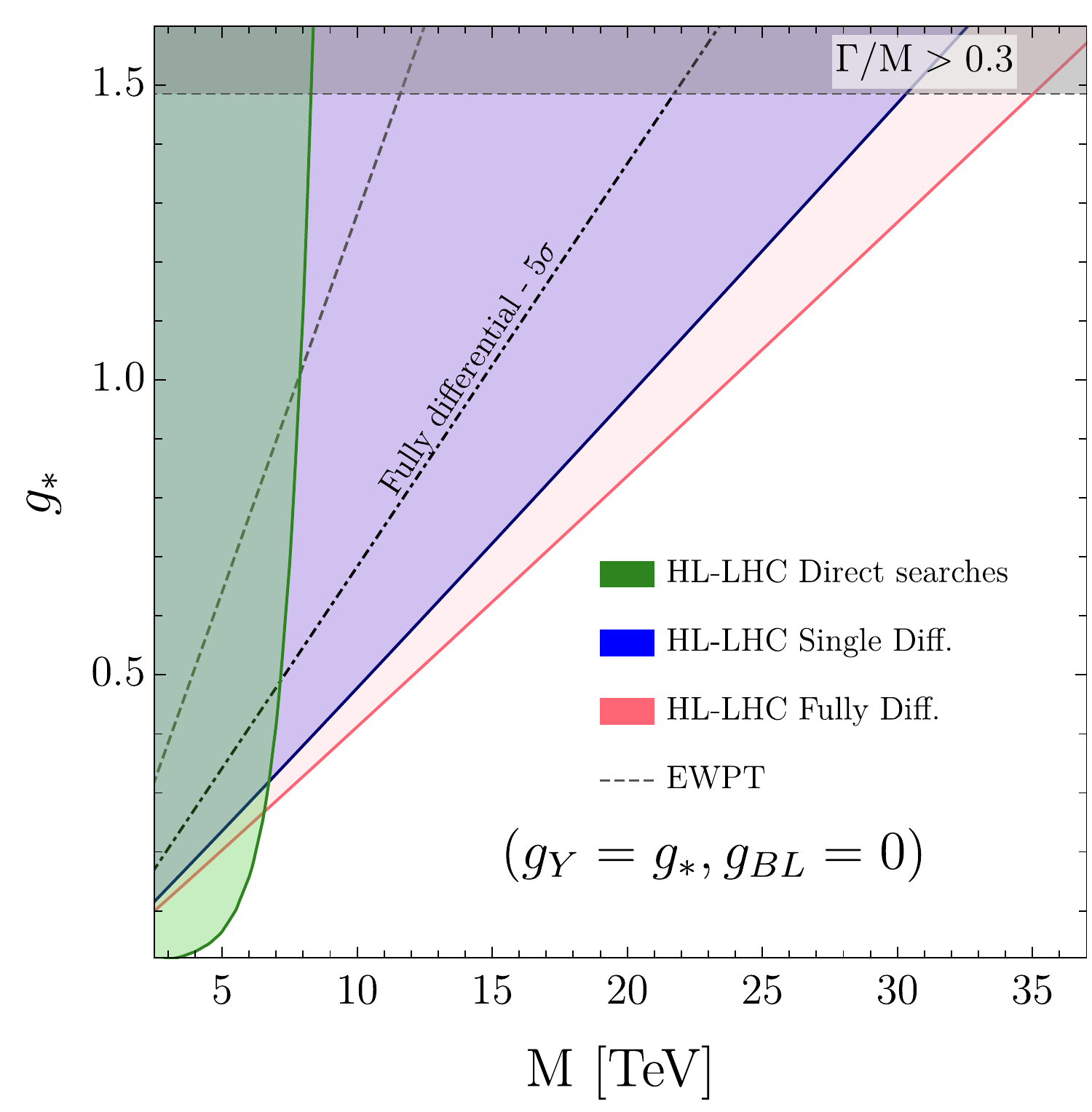}
    \hfill
    \includegraphics[width=0.32\linewidth]{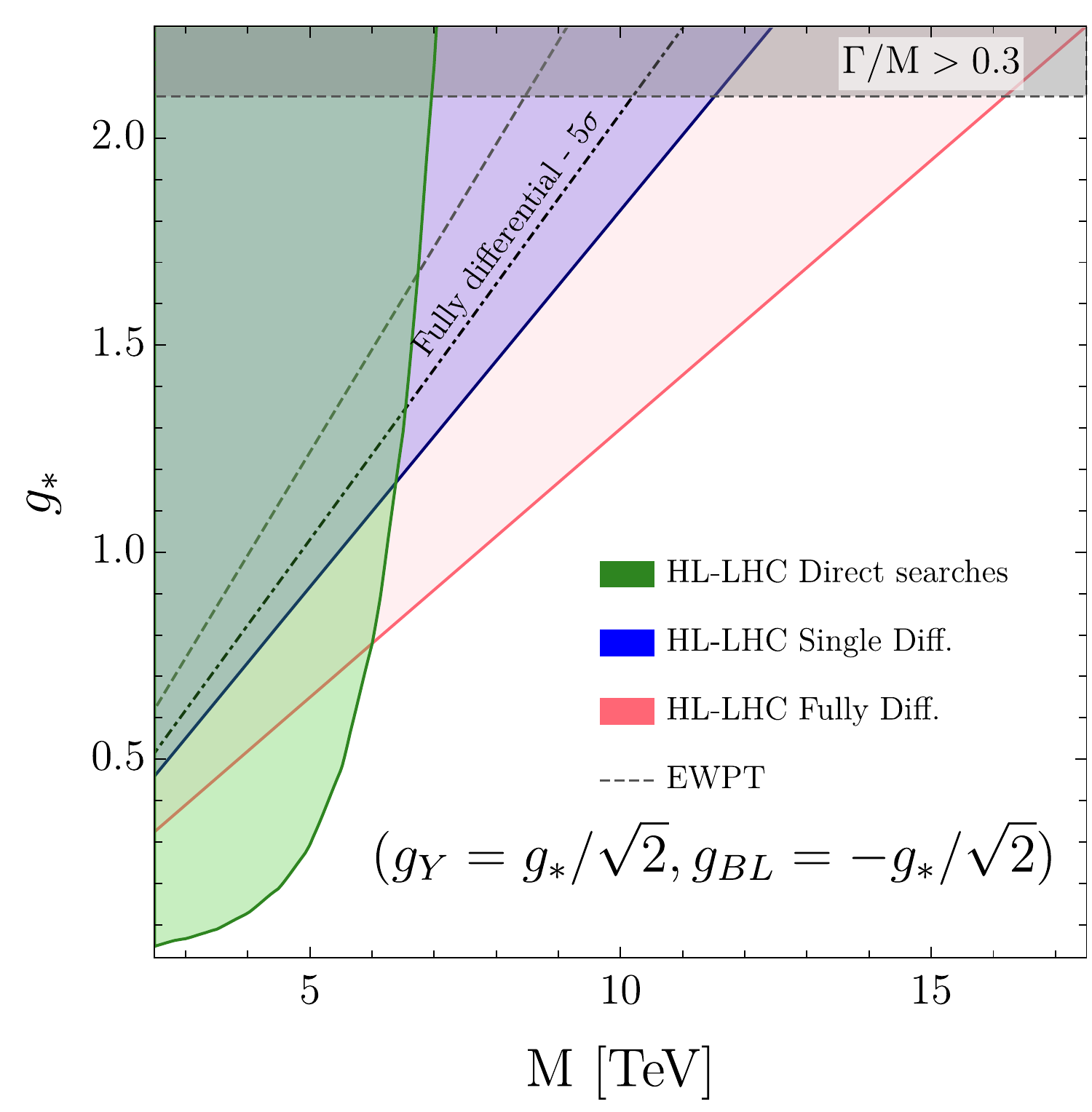}
    \hfill
    \includegraphics[width=0.32\linewidth]{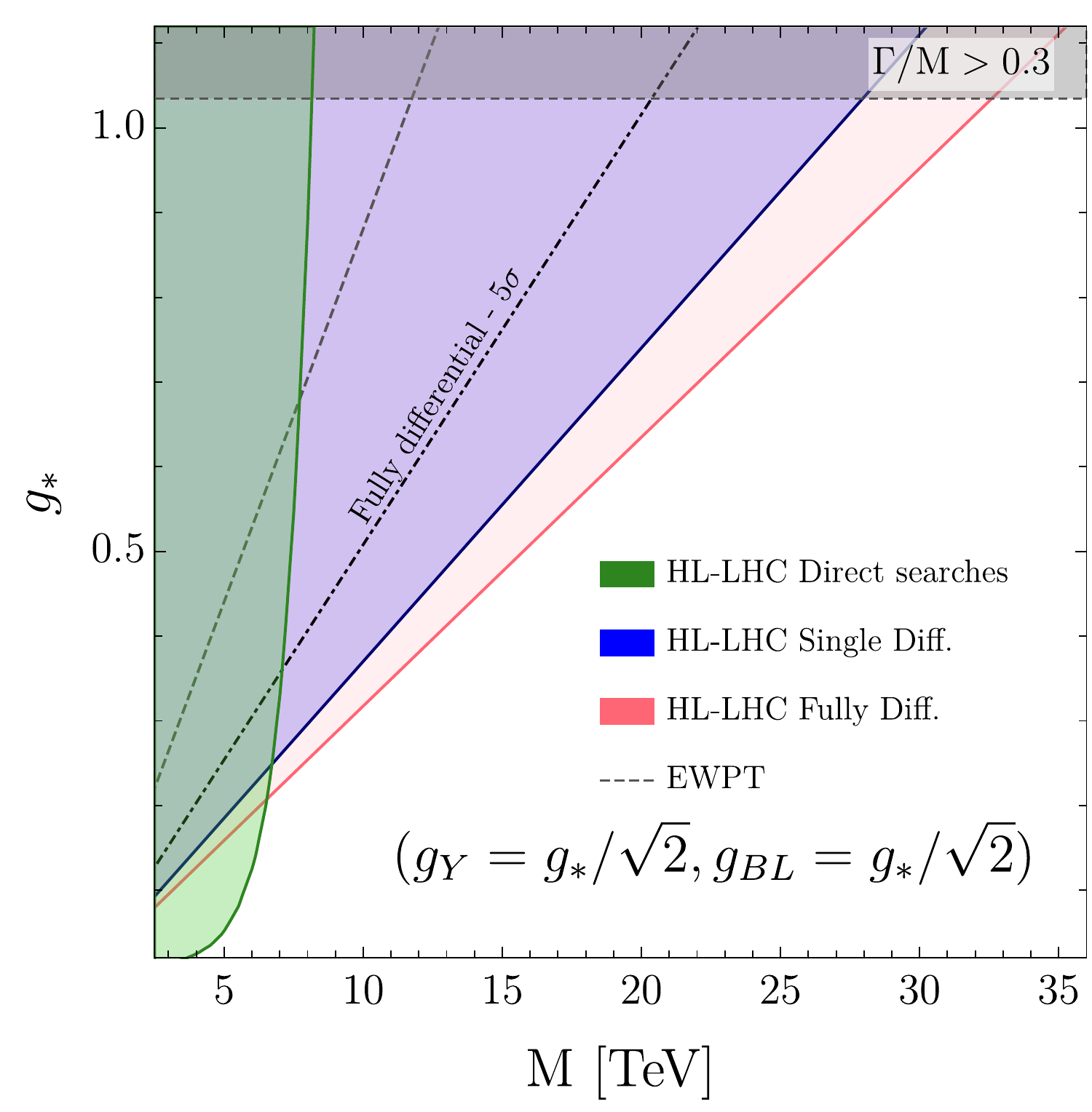}
    \caption{HL-LHC 95{\%} CL (1 d.o.f) exclusion reach in the mass/coupling plane for three different $Z'$ benchmark models, namely $g_{BL} = 0$, $g_{BL} = - g_Y$ and $g_{BL} = g_Y$. The blue shaded region can be excluded through the fully-differential di-lepton DY analysis, while the orange one can be probed with the invariant-mass fit. The green shaded region corresponds to the exclusion from direct searches.}\label{fig:Mgplane}
\end{figure}

When integrated out at tree-level, the massive $Z'$ produces all the flavor-universal lepton-quark operators in Table~\ref{tab:GenOp} except ${\mathcal{O}}_{lq}^{(3)}$, with Wilson coefficients that are readily computed in terms of the three free parameters $g_Y$, $g_{BL}$ and $M$. Clearly the Wilson coefficients are quadratic polynomials in the ratios $g_Y/M$ and $g_{BL}/M$, which are therefore the only two parameter combinations that can be probed by indirect searches\footnote{Strictly speaking this is statement is true only a tree level. In fact, already at our level of accuracy (NLL), the RG flow induces an additional logarithmic dependence on M in our observables. However these effects are extremely small as can be appreciated in Figure~\ref{fig:Mgplane}, where the indirect searches contours are well approximated by straight lines.}.  Furthermore the indirect constraints are  symmetric under an overall change of sign of the couplings $(g_{Y},g_{BL})\rightarrow(-g_{Y},-g_{BL})$. The $95\%$~CL reach on the model at the HL-LHC is displayed in Figure~\ref{fig:Zprime_comp} on the $(g_{Y}/M, g_{BL}/M)$ plane (left panel) and on the $(g_{Y}, g_{BL})$ plane for a fixed mass $M=7$~TeV (right panel). The bounds are obtained from the fully and single-differential analyses described in the previous section, but including in this case a $2\%$ uncorrelated experimental uncertainty in the measurements, aiming at a more conservative result. 

The advantage of the fully-differential analysis over the single-differential one is mainly in the region $g_{BL} \simeq -g_Y$. This region is particularly difficult to probe as it entails the cancellation of the $Z'$ coupling to the right-handed electrons as well as the suppression of the couplings to the left handed quark doublets, the right-handed up-type quarks and the left-handed lepton doublets. Therefore in this region $G_{qe}$, $G_{eu}$ and $G_{ed}$ vanish and $G_{lq}^{(1)}$, $G_{lu}$ and $G_{ld}$ are suppressed (and $\smash{G_{lq}^{(3)}}$ is always zero). We notice in passing that the suppression of the couplings to quarks also determines a reduction of the direct production cross-section at the LHC, which makes direct searches less effective. The fully-differential analysis not only improves the sensitivity along the $g_{BL} = -g_Y$ direction, it also mitigates the impact of the quadratic terms in the cross-section prediction. This is shown by the dashed lines in the left panel of Figure~\ref{fig:Zprime_comp}, reporting the results of the linearized fits. The single-differential linearized analysis possesses two very pronounced flat directions that correspond to directions in the $(g_{Y},g_{BL})$ plane where the $G_{E}^{+}$ coefficient cancels. The fully-differential analysis linearized contour is instead quite close to the full fit thanks to the improved sensitivity to $G_O^+$ at the linear level. 

In the left panel of Figure~\ref{fig:Zprime_comp} we also compare our result with existing EWPT constraints, extracted from Ref.~\cite{Salvioni:2009mt}. With the fully-differential analysis, the progress of the HL-LHC is of a factor around $3$ in $g/M$ in most of the directions in the $(g_{Y}/M, g_{BL}/M)$ plane, which corresponds to an improvement of one order of magnitude in the sensitivity to the Wilson coefficients that scale like $(g/M)^2$. Furthermore, notice that the EWPT bounds in the figure are based on actual experimental measurements whose central value, while compatible with the SM, disfavors the $Z'$ model. This is easily verified in the direction $g_{BL}=0$, where integrating out the $Z'$ produces only the ${\cal O}'_{2B}$ operator with negative coefficient, that corresponds to a positive Y parameter. The central value of Y measured at LEP is instead negative (see e.g.~Ref.~\cite{Farina:2016rws}) making the EWPT exclusion on the model stronger. Our HL-LHC projections assume instead a central value at the SM point. Depending on the sign of central value that will be eventually observed the actual sensitivity to the model could be stronger or weaker than the projection.

We turn now to the comparison of our findings with the projected HL-LHC sensitivity for direct searches of the $Z'$ particle, which are most effectively performed in the dilepton final state. The exclusion on the resonant production cross-section times branching ratio is obtained from the projections in Ref.~\cite{Thamm:2015zwa}, slightly improved to take into account the more recent and refined results in Ref.~\cite{CidVidal:2018eel}. The $Z'$ production cross-section is obtained by two {\sc{MadGraph}} \cite{Alwall:2014hca} simulations (at each $Z'$ mass) with the $Z'$ coupling only to up- or to down-type quarks, rescaled based on the analytical calculation of these couplings as a function of $g_{Y}$ and $g_{BL}$. The branching ratio is also computed analytically. The results are reported in Figure~\ref{fig:Mgplane}, in the mass/coupling plane for three benchmark models $(g_Y=g_*, \, g_{BL} = 0)$, $(g_Y=g_*/\sqrt{2}, \, g_{BL} = -g_*/\sqrt{2})$ and $(g_Y=g_*/\sqrt{2}, \, g_{BL} = g_*/\sqrt{2})$. Notice that the plot extends up to the maximal $g_*$ coupling for which, depending on the model, the width over mass ratio $\Gamma/M$ of the $Z'$ is reasonably small ($<0.3$) enabling a perturbative treatment. The indirect reach from our analyses, and from EWPT, is also reported in the plots. We find a substantial improvement of the mass reach for relatively large $g_*$, up to around $30$~TeV in the first and in the second benchmark model. Finally, in the right panel of Figure~\ref{fig:Zprime_comp} we compare direct and indirect searches in the $(g_{Y}, g_{BL})$ plane at a fixed mass $M=7$~TeV, slightly below the threshold of around $8$~TeV after which direct searches become ineffective. The direction $g_{BL} = -g_Y$ is difficult to probe also directly, as anticipated. The sensitivity improvement of the fully-differential analysis along this direction is significant.

\subsubsection*{Discovery and characterization}

\begin{figure}[t]
\centering
\raisebox{1em}{\includegraphics[width=.54\linewidth]{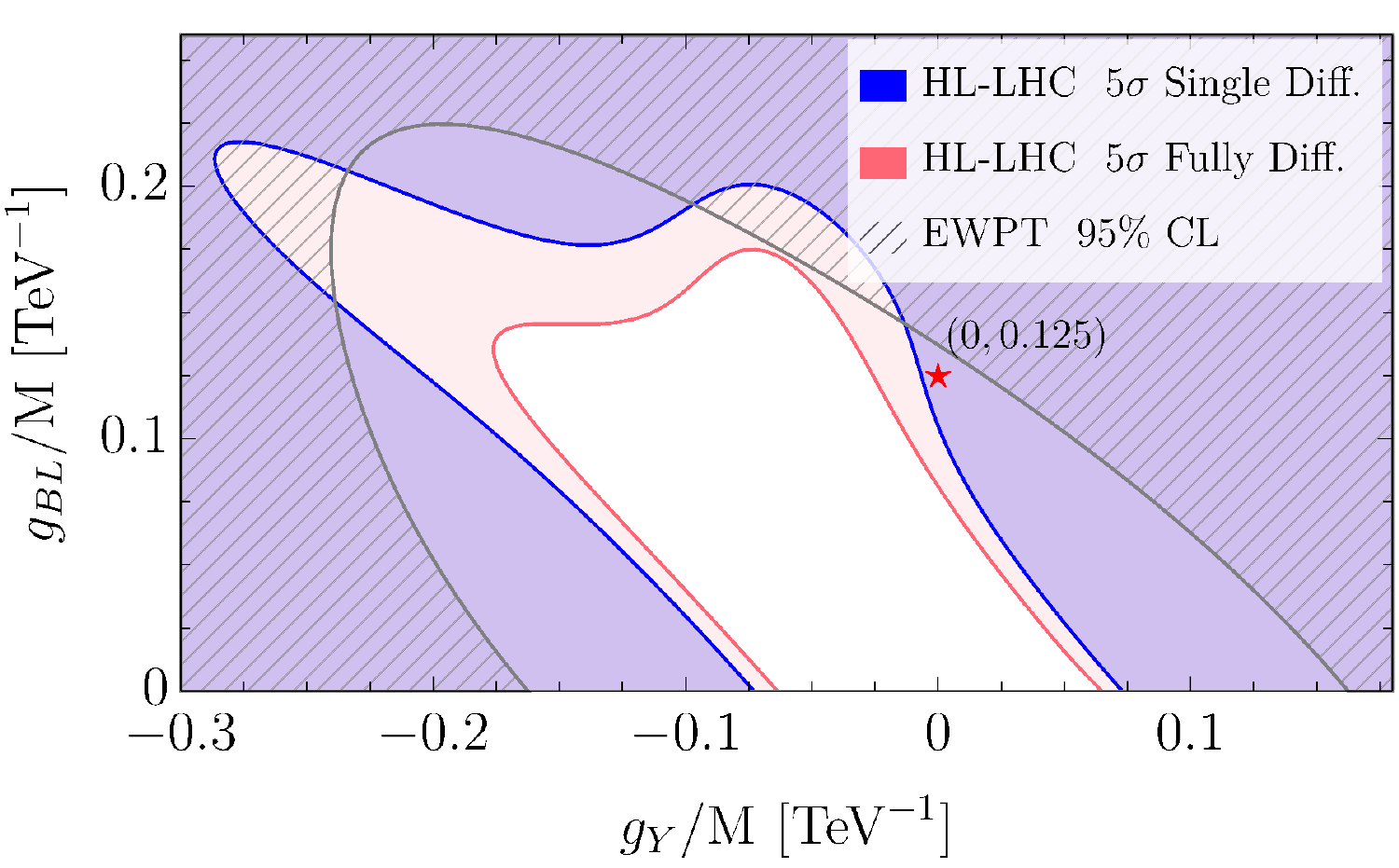}}
\hfill
\includegraphics[width=0.4\linewidth]{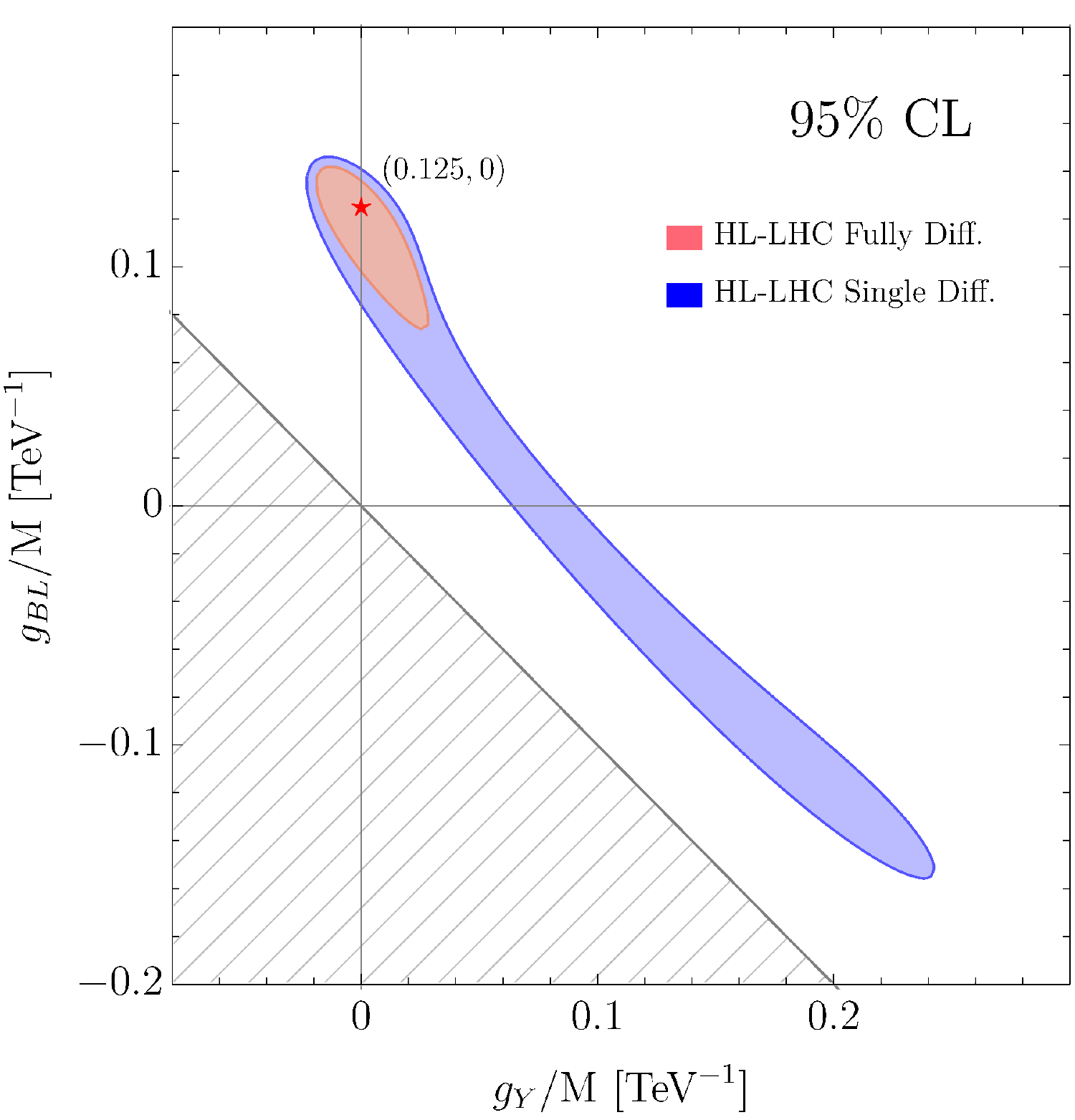}
\caption{Left panel: $5\sigma$ discovery reach for HL-LHC and $95{\%}$ CL EWPT exclusions. Right panel: $95{\%}$ CL likelihood contours assuming the presence of a $Z'$ with $g_{BL}/M=0.12 \;\text{TeV}^{-1}$ in the measurements.}
\label{fig:discovery}
\end{figure}

High-energy measurements have the potential to discover the $Z'$. This is shown on the left panel of Figure~\ref{fig:discovery} by comparing the HL-LHC $5\sigma$~discovery reach with the current exclusion bound from EWPT in the $(g_{Y}/M, g_{BL}/M)$ plane. For $M$ of several TeV or more, direct searches are ineffective and high-energy measurements will provide the only evidence for the existence of the $Z'$. While ``indirect'', i.e.~not based on the detection of a resonant peak, this evidence would be a conclusive and convincing proof of the existence of new physics thanks to the peculiar behavior (growing with energy) of the observed signal and to the possibility of getting confirmations on its nature by the study of angular distributions. The fully-differential analysis would clearly play a major role in this context, on top of course of enabling the discovery itself in a larger region of the parameter space. 

We illustrate the benefits of the fully-differential analysis for the characterization of a putative signal by picking up a point ($g_{BL}/M = 0.12\;\textrm{TeV}^{-1}$ and $g_Y=0$) which is discoverable at the HL-LHC, but close enough to the boundary of the discovery region to make characterization more difficult. We assume the presence of the corresponding signal in the data and we obtain the $95\%$~CL likelihood contours on the right panel of Figure~\ref{fig:discovery}. A simple question related to characterization is whether we can establish that the underlying $Z'$ couples to the $B-L$ current, rather than for instance to the hypercharge current. The figure shows that this is possible only with the fully-differential analysis.

\section{Conclusions}\label{sec:conc}

We studied the potential of fully-differential DY measurements to probe the seven flavor-universal current-current operators listed in Table~\ref{tab:GenOp}. We found (see Table~\ref{tab:SBCC}) that five directions in this parameter space can be probed effectively at the linear interference level with the fully-differential analysis, while with single-differential measurements this is possible for only two directions. At the HL-LHC, the strongest single-parameter sensitivity improvement, by a factor of $3$, is for the parameter ${G}_{O}^{+}$ because of the reasons explained in Section~\ref{treedits}. Improvements in the ballpark of $30\%$ or $10\%$ are observed for the other single-parameter bounds. The improvement is significantly more pronounced for the profiled bounds, owing to the reduction of correlations in the fully-differential fit. The augmented sensitivity at the interference level makes the fully-differential results generically more stable when the quadratic new physics terms are excluded from the predictions. This is beneficial for considerations related to the validity of the EFT, as it lowers the scale of the measurements that drive the sensitivity and reduces the impact of removing the highest-energy cross-section bins as shown in Figure~\ref{fig:Ladder}.

The observed sensitivity improvement is due to two distinct factors. The one with the strongest impact is the extended linear-level sensitivity mentioned above and explained in Section~\ref{treedits} in details. The second, which is also quite a strong effect, is the reduction of the impact of PDF uncertainties due to their correlations across different analysis bins. These correlations are typically different from the ones of the EFT differential cross-section predictions, making harder for the PDF nuisance parameters to mimic the signal and to reduce the sensitivity. Of course the effect is quantitatively so important because the PDF are among the dominant sources of uncertainties in our fit, compatibly with the findings of Ref.~\cite{Ricci:2020xre}. Correspondingly, the benefits of the fully-differential analysis are (mildly) reduced when other sources of systematic uncertainties are assumed to be present, lowering the relative impact of PDF uncertainties in the total error budget. We have verified this fact in Section~\ref{sec:uncertainty} by including a $2\%$ systematic uncertainty uncorrelated across all bins, on top of the fully correlated luminosity uncertainty that is present (but has a totally negligible impact) in the results of Sections~\ref{WYSubSec} and~\ref{GCCSubSec}. 

The dependence (at the $10\%$ or $20\%$ level) of our results on the assumed patterns of experimental systematic uncertainties outlines the need of detailed experimental projections for DY measurements. Experimental uncertainties that are fully correlated in all bins as in Sections~\ref{WYSubSec} and~\ref{GCCSubSec} are definitely unrealistic. However assuming the uncertainties to be fully uncorrelated, as in Section~\ref{sec:uncertainty}, is equally unrealistic. We do expect correlations, especially in the fully-differential measurements, whose impact could be beneficially for the sensitivity analogously to what we have found happening for the PDF. The final HL-LHC sensitivity could thus be closer to the one in Sections~\ref{WYSubSec} and~\ref{GCCSubSec} than to the one in Section~\ref{sec:uncertainty}. Furthermore, our findings are based on the statistically sub-optimal strategy of comparing cross-section measurements with EFT predictions, rather than comparing directly the EFT with the observed data. More sophisticated and unbinned strategies could be considered to further improve the sensitivity.

A significant improvement in the sensitivity is also found in the LHC run 3 projections (see Appendix~\ref{app:LHCProj}). The gain is however much milder than for the HL-LHC, mostly because the number of expected events is too low to efficiently reconstruct the full angular distributions at high energy (i.e.~to populate enough all the bins required for a fully-differential analysis).

The sensitivity improvement of fully-differential measurements has a direct impact on concrete putative new physics scenarios, as we discussed in Section~\ref{sec:Zprime} for a simple minimal $Z^\prime$ model. The point is that in models where the charged current $\smash{{\mathcal{O}}_{lq}^{(3)}}$ operator is absent, the single-differential DY analysis is mostly sensitive to a single EFT parameters combination: $\smash{G_{E}^{+}}$. In the new physics model it will be generically possible to suppress $\smash{G_{E}^{+}}$ without particular fine-tuning, making the single-differential analysis loose sensitivity in a large region of the parameter space as in Figure~\ref{fig:Zprime_comp}. The fully-differential analysis will boost the sensitivity in that region. 

The results of Section~\ref{sec:Zprime} also outline the effectiveness of high-$p_T$ probes on a well-established new physics benchmark that has been investigated since the beginning of the LHC program. High-$p_T$ probes by high energy DY measurements extend (see Figure~\ref{fig:Mgplane}) the projected HL-LHC exclusions well beyond the reach of direct searches in a large region of the parameter space, with sensitivity to masses from $10$ to $30$~TeV. Discovery is also possible up to around $20$~TeV. In the event of a discovery, fully-differential measurements will play a crucial role in the characterization of the observed signal as illustrated in Figure~\ref{fig:discovery}.

\section*{Acknowledgements}
The work of L.R. was supported by the Swiss National Science Foundation under contract 200021-178999. G.P.~was supported in part by the MIUR under contract 2017FMJFMW (PRIN2017).

\appendix

\section{Kinematical variables}\label{app:Kin}

The charged leptons momenta we employ in our analyses are obtained by recombining photons (and lepton pairs from photon splitting) within a $\Delta R^{\rm{rec}}=0.1$ recombination cone, with thresholds $p_{T,{\rm{min}}}^\gamma=10$~GeV and $|\eta|_{\rm{max}}^\gamma=3$. Acceptance cuts $p_{T,{\rm{min}}}^\ell=25$~GeV and $|\eta|_{\rm{max}}^\ell=2.5$ are applied to the reconstructed lepton momenta. Events are selected to have $2$ (same-flavor) or $1$ reconstructed leptons for, respectively, the neutral and charged DY analyses. 

The two variables employed in the charged analysis are the transverse momentum $p_{T,\ell}$ and the rapidity $\eta_\ell$ of the single observed lepton. 

In the neutral case, the three variables are the invariant mass $m_{\ell\ell}$ of the dilepton pair, the rapidity of the dilepton system relative to the beam axis in absolute value (called $y$) and the cosine of the scattering angle $c_*=\cos\theta_*$. Two alternative definitions of $\theta_*$ can be considered. The simplest option is to define it as the angle formed, in the rest frame of the dilepton pair, between the charge-minus lepton and the direction of motion of the dilepton rest frame relative to the lab frame. The second option is to define $\theta_*$ as the angle between the momentum of the charge-minus lepton and the $z$ axis in the Collins--Soper frame, along the lines of Ref.~\cite{Collins:1977iv}. More precisely, $c_*$ is defined by the formula
\begin{equation}\displaystyle\label{CSan}
\displaystyle c_* = \frac{2 p_{z,\ell\ell}}{m_{\ell\ell} |p_{z,\ell\ell}|} \frac{p_{z,\ell^+} E_{\ell^-} - p_{z,\ell^-} E_{\ell^+}}{(m_{\ell\ell}^2 + p_{\bot, \ell\ell}^2)^{1/2}}\,,
\end{equation}
where $E_{\ell^\pm}$ and $p_{z,\ell^\pm}$ denote the energy and longitudinal momentum of each lepton, while $p_{z,\ell\ell}$ and $p_{\bot,\ell\ell}$ are the longitudinal and transverse momentum of the lepton pair. The two definitions only coincide for the tree-level kinematics, where the dilepton pair has vanishing transverse momentum. However we have checked that the differential distributions obtained with the two definitions are almost identical, thus giving nearly equal results for the fits (the differences in the bounds being at most $2-3\%$). The results reported in the paper are obtained with the Collins--Soper definition of $c_*$.

We remark that the definition in eq.~(\ref{CSan}) contains the factor $p_{z,\ell\ell}/|p_{z,\ell\ell}|$, which takes into account the direction of the boost of the CoM system. Its presence is essential for the effectiveness of the fully-differential analysis as emphasized in Footnote~\ref{foot1}. This factor is duly taken into account in existing experimental measurements (see e.g.~Ref.~\cite{Aaboud:2017ffb}).

\section{Explicit formulas}\label{app:Form}

In this appendix we collect the explicit expressions for the tree-level amplitudes.

Adopting the notation of Ref.~\cite{Ricci:2020xre}, the four coefficient functions $P^q_{s,o}$ appearing in the fully-differential cross-section for the neutral DY process, eq.~\eqref{eq:sigma_ll}, can be expressed as
\begin{align}
    P_{s}^q = \hat{s}^2\left(\left( {\mathcal{C}}_0(q_{L},\,l_{L}) \right)^2+\left( {\mathcal{C}}_0(q_{R},\,l_{R}) \right)^2\right) \,,&&  P_{o}^q =\hat{s}^2\left(\left( {\mathcal{C}}_0(q_{L},\,l_{R}) \right)^2+\left( {\mathcal{C}}_0(q_{R},\,l_{L}) \right)^2\right)\,,
\end{align}
where the $C_0$ functions can be split as
\begin{equation}\displaystyle\label{eq:C0}
 {\mathcal{C}}_0(q_{\chi_q},\,l_{\chi_l})={\mathcal{C}}^0_{\rm{SM}}(s;q_{\chi_q},\,l_{\chi_l})+{\cal{K}}^0_{{{q_{\chi_q},\,l_{\chi_l}}}}\,,
\end{equation}
with a SM contribution
\begin{equation}\displaystyle
{\mathcal{C}}^0_{\rm{SM}}(\hat{s};q_{\chi_q},\,l_{\chi_l})=\frac{g^2 T^3(q_{\chi_q}) T^3(l_{\chi_l})}{\hat{s}} + \mathcal{O}\left(\frac{m_Z^2}{\hat{s}} \right)\,,
\end{equation}
and new physics effects encoded in ${\cal{K}}^0_{{{q_{\chi_q},\,l_{\chi_l}}}}$. The explicit expressions for the latter are reported in Table~\ref{tab:GenK} for the quark-lepton interaction in the Warsaw basis, and in Table~\ref{tab:WYK} for the $W$ and $Y$ parameters. 

\begin{table}[t]
\centering
\renewcommand{\arraystretch}{1.5}
\begin{tabular}{c|cccc}
($\chi_q$,$\chi_l$)&(L,L)&(L,R)&(R,L)&(R,R)\\
\hline
\rule{0pt}{1.5em}$u$ -- $\bar u$ subprocess &$G_{lq}^{(1)}-G_{lq}^{(3)}$&$G_{qe}$&$G_{lu}$&$G_{eu}$\\
\rule{0pt}{1.5em}$d$ -- $\bar d$ subprocess &$G_{lq}^{(1)}+G_{lq}^{(3)}$&$G_{qe}$&$G_{ld}$&$G_{ed}$
\end{tabular}
\caption{Explicit expressions for the new physics coefficients ${\cal{K}}^0_{{{q_{\chi_q},\,l_{\chi_l}}}}$ defined in eq.~(\ref{eq:C0}) for the generic quark-lepton operators in the Warsaw basis (Table~\ref{tab:GenOp}).}\label{tab:GenK}
\centering
\renewcommand{\arraystretch}{1.5}
\begin{tabular}{c|cccc}
($\chi_q$,$\chi_l$)&(L,L)&(L,R)&(R,L)&(R,R)\\
\hline
\rule{0pt}{1.5em}$u$ -- $\bar u$ subprocess &$\frac{3 g^2 \text{W}+g'^2 \text{Y}}{12 m_W^2}$&$\frac{g'^2 Y}{6 m_W^2}$&$\frac{g'^2 \text{Y}}{3 m_W^2}$&$\frac{2 g'^2 \text{Y}}{3 m_W^2}$\\
\rule{0pt}{1.75em}$d$ -- $\bar d$ subprocess &$\frac{-3 g^2 \text{W}+g'^2 \text{Y}}{12 m_W^2}$&$\frac{g'^2 \text{Y}}{6 m_W^2}$&$-\frac{g'^2 \text{Y}}{6 m_W^2}$&$-\frac{g'^2 \text{Y}}{3 m_W^2}$
\end{tabular}
\caption{Explicit expressions for the  new physics coefficients ${\cal{K}}^0_{{{q_{\chi_q},\,l_{\chi_l}}}}$ defined in eq.~(\ref{eq:C0}) for the $\mathcal{O}_{2W}'$ and $\mathcal{O}_{2B}'$ operators.}\label{tab:WYK}
\end{table}

From the previous equations one can easily derive the expressions for the vectors $\vec{V}_{s,o}^{q}$ defined in eq.~(\ref{P}) as
\begin{equation}
G_{s,o}^{u,d} \equiv \vec V_{s,o}^{u,d} \cdot \vec G\,,
\end{equation}
where $\vec G$ are the Wilson coefficients in the Warsaw basis. Analogously we defined the $\vec V_{s,o}^\bot$ vectors
\begin{equation}
G_{s,o}^{\bot} \equiv \vec V_{s,o}^{\bot} \cdot \vec G\,.
\end{equation}
We recall that the $G_{s,o}^{\bot}$ coefficients are chosen to have components only along the operators contributing to the neutral DY process and to be orthogonal to the coefficient combinations ${G}_{s,o}^{u,d}$. Moreover $\vec V_{s}^\bot$ and $\vec V_{o}^\bot$ contribute to same-chirality and opposite-chirality subprocesses, respectively.
We report the explicit values of the $\vec V$ components in Table~\ref{tab:VSVD}.

The explicit expressions for the $G_{E}^\pm$ and $G_O^\pm$ coefficients can be derived by substituting the above expressions in the definitions in eqs.~(\ref{eq:GE+def}),~(\ref{eq:GE-def}) and~(\ref{eq:GOdef}).

\begin{table}[t]
\small
\centering
\begin{tabular}{@{\hspace{.25em}}c|ccccccc|cc@{\hspace{.25em}}}
\rule[-.5em]{0pt}{1.em}&\multicolumn{7}{c|}{Generic quark-lepton operators}&\multicolumn{2}{c}{W and Y}\\
\hline
\rule[-.6em]{0pt}{2.em}&${G}_{lq}^{(3)}$&${G}_{lq}^{(1)}$&${G}_{qe}$&${G}_{lu}$&${G}_{ld}$&${G}_{eu}$&${G}_{ed}$&{W}&{Y}\\
\hline
\rule[-.5em]{0pt}{2.25em}$\vec{V}_{s}^u$&$\frac{3 g^2+{g'}^2}{6}$&$\frac{-3 g^2-{g'}^2}{6}$&$0$&$0$&$0$&$-\frac{4 {g'}^2}{3}$&$0$&$-\frac{g^2\left(3 g^2+{g'}^2\right)}{24 m_W^2}$&$-\frac{{g'}^2 \left(3g^2+65 {g'}^2\right)}{72 m_W^2}$\\
\rule[-.5em]{0pt}{2.25em}$\vec{V}_{s}^d$&$\frac{3 g^2-{g'}^2}{6}$&$\frac{3 g^2-{g'}^2}{6}$&$0$&$0$&$0$&$0$&$\frac{2 {g'}^2}{3}$&$-\frac{g^2 \left(3 g^2 - {g'}^2\right)}{24 m_W^2}$&$\frac{g'^2\left(3 g^2 - 17 {g'}^2\right)}{72 m_W^2}$\\
\rule[-.5em]{0pt}{2.25em}$\vec{V}_{o}^u$&$0$&$0$&$-\frac{{g'}^2}{3}$&$-\frac{2{g'}^2}{3}$&$0$&$0$&$0$&$0$&$-\frac{5 {g'}^4}{18 m_W^2}$\\
\rule[-.5em]{0pt}{2.25em}$\vec{V}_{o}^d$&$0$&$0$&$-\frac{{g'}^2}{3}$&$0$&$\frac{{g'}^2}{3}$&$0$&$0$&$0$&$-\frac{{g'}^4}{9 m_W^2}$\\
\rule[-.5em]{0pt}{2.25em}$\vec{V}_{s}^\bot$&$0$&$\frac{2 g'^2}{3}$&$0$&$0$&$0$&$\frac{-3g^2-g'^2}{12}$&$\frac{g'^2-3g^2}{6}$&$0$&$-\frac{{g'}^4}{18 m_W^2}$\\
\rule[-.5em]{0pt}{2.25em}$\vec{V}_{o}^\bot$&$0$&$0$&$\frac{2{g'}^2}{3}$&$-\frac{g'^2}{3}$&$\frac{2{g'}^2}{3}$&$0$&$0$&$0$&$-\frac{{g'}^4}{9 m_W^2}$
\end{tabular}
\caption{Explicit expressions for the components of the $\vec{V}_{s,o}^q$ and $\vec V_{s,o}^\bot$ vectors.
The first part of the table gives the components along the generic quark-lepton operators in the Warsaw basis, while the last two columns correspond to the components along the W and Y parameters.}\label{tab:VSVD}
\end{table}

\section{LHC projections}\label{app:LHCProj}

In this appendix we report the projections for the LHC run 3, assuming an energy of $14\;$TeV and an integrated luminosity $\mathcal{L} = 300\,\text{fb}^{-1}$.

The $95{\%}$ CL, single-parameter bounds for the Universal parameters W and Y are given by
\begin{equation}\displaystyle\label{eq:boundsWY_LHC}
\begin{array}{rll}
\textrm{W}: &\quad [-6.8, 7.2] \times 10^{-5} \quad & ([-8.2, 8.9] \times 10^{-5})\,,\\
\rule{0pt}{1.5em}\textrm{Y}: & \quad [-17,19] \times 10^{-5} \quad & ([-18, 20] \times 10^{-5})\,,
\end{array}
\end{equation}
where the two sets of intervals correspond to the fully-differential and single-differential combined bounds (the latter in brackets).
These bounds are obtained including only a $2\%$ luminosity uncertainty, but no additional experimental systematic error.
Comparing with the analogous HL-LHC bounds in eq.~\eqref{eq:boundsWY}, we see that the all bounds are roughly a factor $2$ weaker. The fully-differential analysis provides an improvement in the sensitivity, although significantly milder than in the HL-LHC case.
This pattern is not unexpected, since the benefits of a fully-differential analysis tend to be larger when the number of expected events is large enough to allow the fits to be dominated by the linear interference terms. At the LHC with $\mathcal{L} = 300\,\text{fb}^{-1}$ the importance of the quadratic terms in the fits is still high, thus explaining the much milder gain in sensitivity. As we will see in the following, a similar pattern is found for the fits on the seven 4-fermion operators and on the minimal $Z'$ model.

In Tables~\ref{tab:SBCCLHC} and~\ref{tab:Bounds_WarsawLHC} we report the LHC run 3 bounds for the single parameter and the profiled fits in the basis introduced in Section~\ref{GCCSubSec} and in the Warsaw basis, respectively. Also in this case the bounds are derived assuming no uncorrelated experimental systematic uncertainty. As for the HL-LHC, the main benefit of the fully-differential fit is in the determination of the $G_O^+$ parameter, which improves by roughly $40\%$ in the single-parameter fit. Mild improvements in the sensitivity to the other parameters (of order $10 - 20\%$) are also found.

\begin{table}[t]
    \centering
    {\small
    \begin{tabular}{@{\hspace{.2em}}c@{\hspace{.2em}}|c@{\hspace{.65em}}c@{\hspace{.65em}}c|c@{\hspace{.65em}}c@{\hspace{.65em}}c@{\hspace{.2em}}}
    {\bf{95\%CL}}&\multicolumn{3}{c|}{single parameter}&\multicolumn{3}{c}{profiled}\\
    \rule[-.6em]{0pt}{1.85em}$[10^{-3}\,\textrm{TeV}^{-2}]$ & fully diff. & fully diff. lin. & {\rm{single diff.}} & fully diff. & fully diff. lin. & single diff.\\
      \hline
    \rule{0pt}{1.35em}$\widetilde{G}_{lq}^{(3)}$ & $[-0.87 ,\, 0.87]$ & $[-0.87,\, 0.87]$ & $[-0.97 ,\, 0.97]$
    & $[-1.19 ,\, 1.06]$ & $[-1.30,\, 1.30]$ & $[-1.51 ,\, 1.27]$\\
    \rule{0pt}{1.35em}${G}_{E}^{+}$ & $[ -0.53 ,\, 0.47]$ & $[-0.50,\, 0.50]$ &  $[-0.60 ,\, 0.53]$
        & $[-1.30 ,\, 0.74]$ & $[-1.50,\, 1.50]$ & $[-2.01 ,\, 0.85]$\\
    \rule{0pt}{1.35em}${G}_{E}^{-}$ & $[-1.13 ,\, 1.20]$ & $[-2.66,\, 2.66]$ & $[-1.30 ,\, 1.51]$
        & $[-3.56 ,\, 2.80]$ & $[-4.54,\, 4.54]$ & $[-5.03 ,\, 3.29]$\\
    \rule{0pt}{1.35em}${G}_{O}^{+}$ &  $[-1.07,\, 0.87]$ & $[-1.00,\, 1.00]$ & $[-1.74 ,\, 1.94]$
        & $[-1.16 ,\, 1.45]$ & $[-2.17,\, 2.17]$ & $[-2.16,\, 3.41]$\\
    \rule{0pt}{1.35em}${G}_{O}^{-}$ & $[-1.30,\, 1.16]$ & $[-7.88,\, 7.88]$ & $[-1.53 ,\, 1.35]$
        & $[-3.02,\, 3.70]$ & $[-17.9,\, 17.9]$ & $[-3.41 ,\, 4.90]$\\
    \rule{0pt}{1.35em}${G}_s^{\bot}$ & $[-3.48,\, 2.75]$ & $[-11.0,\, 11.0]$ & $[-3.52 ,\, 2.81]$
        & $[-4.05 ,\, 3.79]$ & $[-30.5,\, 30.5]$ & $[-4.38,\, 3.92]$\\
     \rule[-.75em]{0pt}{2.1em}${G}_o^{\bot}$ & $[-1.19,\, 1.17]$ & $[-51.9,\, 51.9]$ & $[-1.42 ,\, 1.40]$
        & $[-1.65,\, 1.47]$ & $[-70.2,\, 70.2]$ & $[-2.00,\, 2.00]$
    \end{tabular}
   }
       \caption{$95{\%}$ CL projected bounds (in $10^{-3}\;\textrm{TeV}^{-2}$ units) for the seven parameters $\widetilde{G}_{lq}^{(3)}$, $G_{E,O}^\pm$ and $G_{s,o}^\bot$ at the $14\;$TeV LHC with $\mathcal{L} = 300\,\text{fb}^{-1}$.}
        \label{tab:SBCCLHC}
    \end{table}
\begin{table}[t]
    \centering
    {\small
    \begin{tabular}{@{\hspace{.2em}}c@{\hspace{.2em}}|c@{\hspace{.65em}}c@{\hspace{.65em}}c|c@{\hspace{.65em}}c@{\hspace{.65em}}c@{\hspace{.2em}}}
     {\bf{95\%CL}}&\multicolumn{3}{c|}{single parameter}&\multicolumn{3}{c}{profiled}\\
    \rule[-.6em]{0pt}{1.85em}$[10^{-3}\,\textrm{TeV}^{-2}]$ & fully diff. & fully diff. lin. & single diff. & fully diff. & fully diff. lin. & single diff.\\
    \hline
    \rule{0pt}{1.35em}${G}_{lq}^{(3)}$ & $[-1.09,\, 1.03]$ & $[-1.06,\, 1.06]$ & $[-1.35,\, 1.25]$
    & $[-1.19,\, 1.06]$ & $[-1.30,\, 1.30]$ & $[-1.51,\, 1.27]$\\
    \rule{0pt}{1.35em}${G}_{lq}^{(1)}$ & $[-4.68,\, 7.51]$ & $[-5.66,\, 5.66]$ & $[-5.11,\, 10.64]$
        & $[-8.62,\, 10.56]$ & $[-35.6,\, 35.6]$ & $[-9.72,\, 13.5]$\\
    \rule{0pt}{1.35em}${G}_{qe}$ & $[-4.71,\, 8.62]$ & $[-6.57,\, 6.57]$ & $[-6.11,\, 10.9]$
        & $[-7.23,\, 9.37]$ & $[-375,\, 375]$ & $[-10.9,\, 11.3]$\\
    \rule{0pt}{1.35em}${G}_{lu}$ & $[-4.22,\, 6.42]$ & $[-5.01,\, 5.01]$ & $[-5.88,\, 10.6]$
        & $[-6.05,\, 10.3]$ & $[-187,\, 187]$ & $[-11.6,\, 13.7]$\\
    \rule{0pt}{1.35em}${G}_{ld}$ & $[-16.3,\, 10.2]$ & $[-18.0,\, 18.0]$ & $[-19.2,\, 12.5]$
        & $[-16.9,\, 14.9]$ & $[-419,\, 419]$ & $[-20.2,\, 20.9]$\\
    \rule{0pt}{1.35em}${G}_{eu}$ & $[-3.10,\, 3.54]$ & $[-3.30,\, 3.30]$ & $[-3.29,\, 3.83]$
        & $[-6.20,\, 10.3]$ & $[-46.9,\, 46.9]$ & $[-7.25,\,11.7]$\\
    \rule[-.75em]{0pt}{2.1em}${G}_{ed}$ & $[-13.5,\, 8.18]$ & $[-10.3,\, 10.3]$ & $[-15.5,\, 8.69]$
        & $[-16.9,\, 16.0]$ & $[-121,\, 121]$ & $[-17.9,\, 18.7]$
    \end{tabular}}
     \caption{$95{\%}$ CL projected bounds (in $10^{-3}\;\textrm{TeV}^{-2}$ units) for the four-fermion operator coefficients in the Warsaw basis at the $14\;$TeV LHC with $\mathcal{L} = 300\,\text{fb}^{-1}$.
        }
        \label{tab:Bounds_WarsawLHC}
\end{table}    

\begin{figure}[t]
    \centering
    \includegraphics[width=0.32\linewidth]{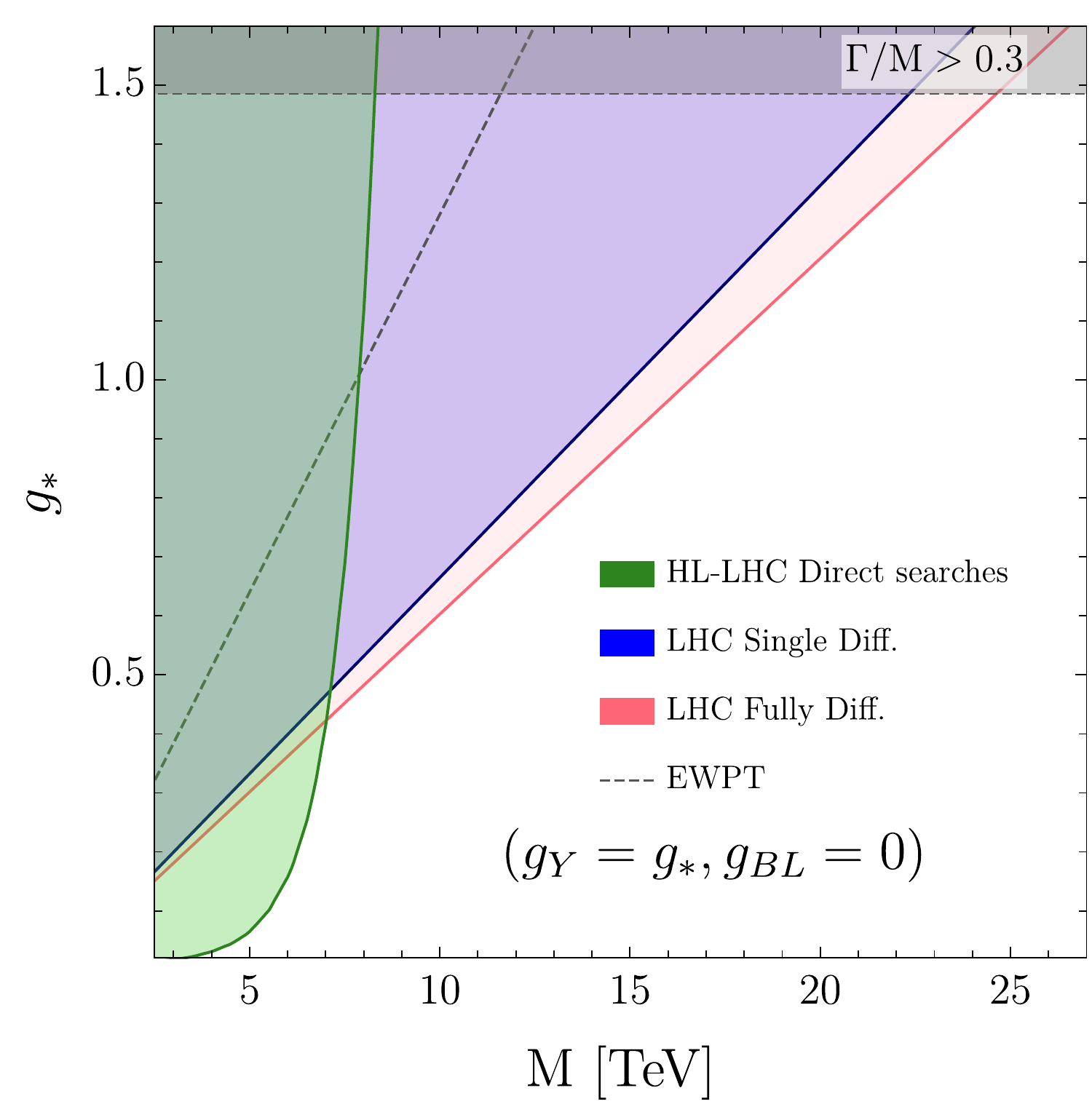}
    \hfill
    \includegraphics[width=0.32\linewidth]{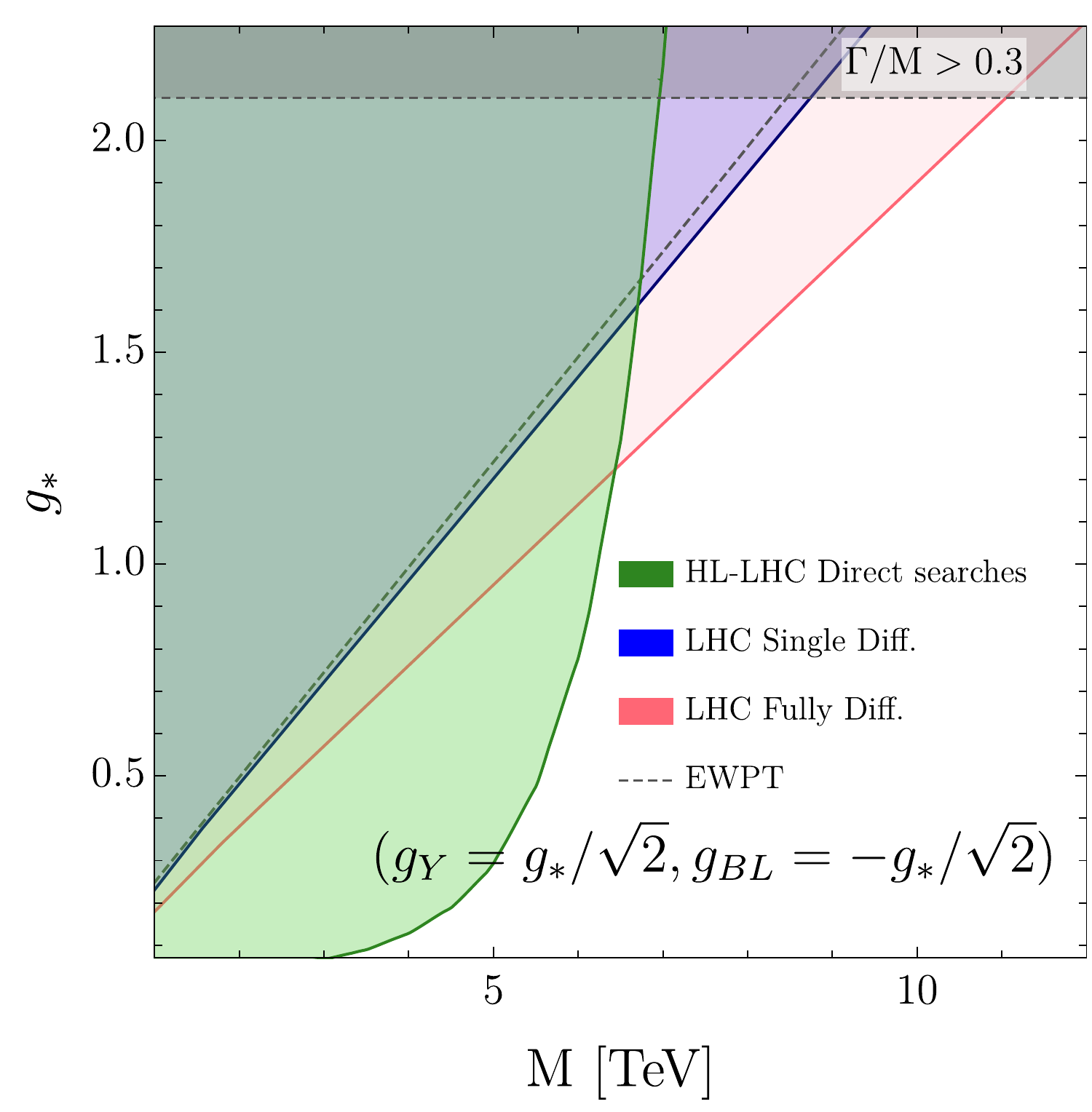}
    \hfill
    \includegraphics[width=0.32\linewidth]{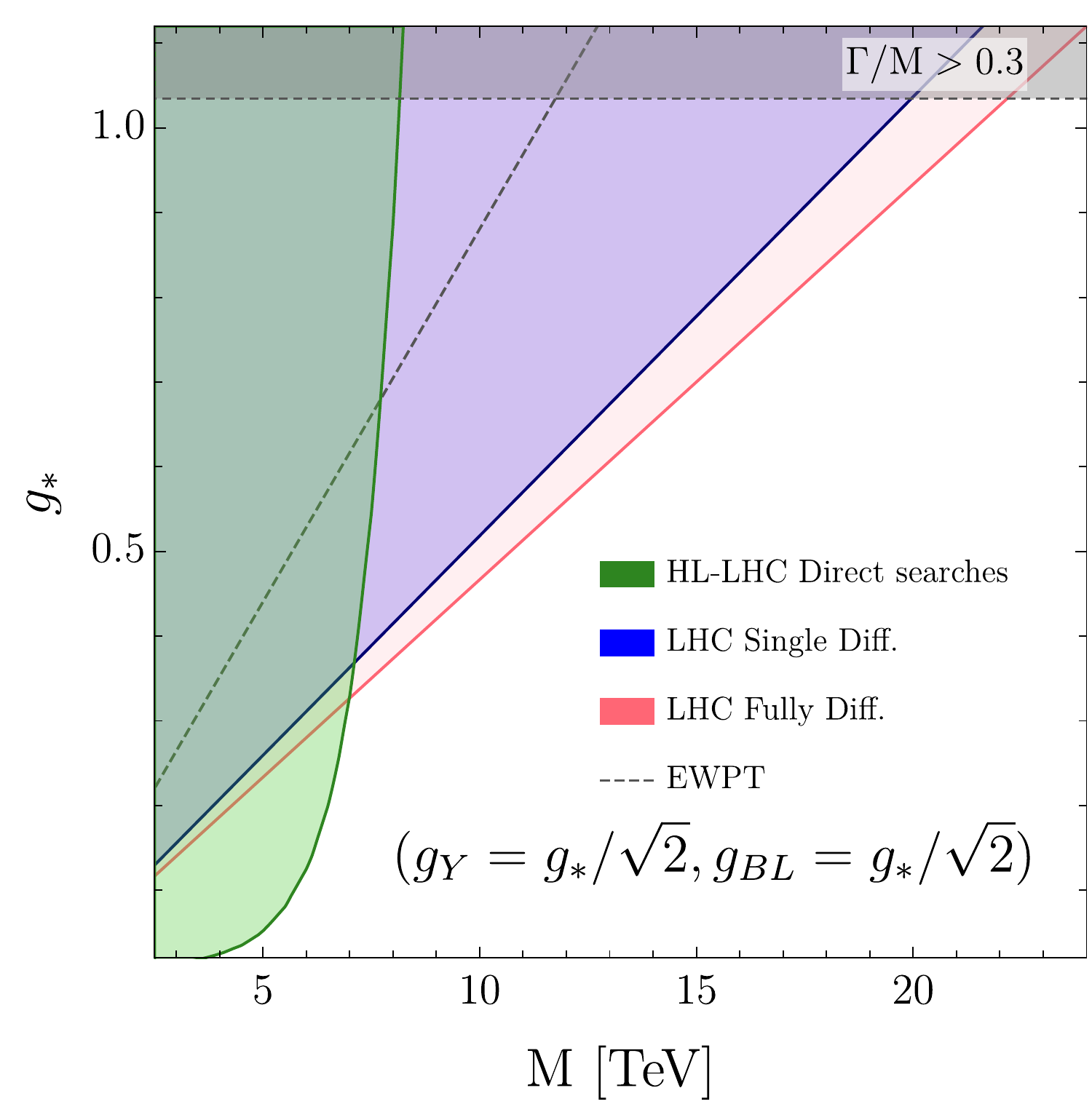}
    \caption{95{\%} CL (1 d.o.f) exclusion reach in the mass/coupling plane for three different $Z'$ benchmark models at LHC with $\mathcal{L} = 300$ $\text{fb}^{-1}$. The blue shaded region can be excluded through the fully-differential di-lepton DY analysis, while the orange one can be probed with the invariant-mass fit. The green shaded region corresponds to the exclusion from direct searches.\label{fig:MgplaneLHC}}
\end{figure}

Finally in Figure~\ref{fig:MgplaneLHC} we report the projected LHC exclusion reach in the mass/coupling plane for the three benchmark $Z^{\prime}$ models considered in Section~\ref{sec:Zprime}. The mass exclusion reach is roughly $25\%$ weaker than the projections for the HL-LHC (compare Figure~\ref{fig:Mgplane}). The improvement due to the fully-differential analysis is clearly visible. In particular it plays a significant role in the difficult-to-test benchmark with $g_Y = - g_{BL}$, allowing one to cover part of the parameter space not yet excluded by the EWPT, which is not accessible with the single-differential analysis.

\providecommand{\href}[2]{#2}\begingroup\raggedright\endgroup

\end{document}